\documentclass[aps,prl,onecolumn,notitlepage]{revtex4-2}
\usepackage[a4paper,top=2.00cm,bottom=2.00cm,left=2.00cm,right=2.00cm]{geometry}
\usepackage{amsmath}
\usepackage{amssymb}
\usepackage{latexsym}
\usepackage{enumerate}
\usepackage{caption}
\usepackage{subcaption}
\usepackage{nicefrac}
\usepackage[utf8]{inputenc}
\fontfamily{cmr}\selectfont
\usepackage{graphicx}
\usepackage{ifpdf}
\usepackage{epstopdf}
\DeclareGraphicsExtensions{.eps, .pdf, .jpg, .png, .tif}
\usepackage[]{hyperref}
\usepackage{float}
\usepackage[force]{feynmp-auto}
\usepackage{bookmath}

\newcommand*{\nf}{n_{\mathrm{F}}}
\newcommand*{\V}[1]{\mathbf{#1}}
\newcommand*{\fm}{\mathrm{F}}

\newcommand*{\refsub}[2]{\hyperref[#1]{\ref{#1}#2}}
\def\sml{{\bf\textsf{s}}}
\def\cO{{\mathcal O}}
\newcommand{\mathleft}{\@fleqntrue\@mathmargin0pt}
\newcommand{\atanfactor}{\arctan\frac{x_\mathrm{c}}{\sqrt{\vartheta}}}
\begin{document}
\title{Effects of Incipient Pairing on Non-equilibrium \\ Quasiparticle Transport in Fermi Liquids}
\author{Wei-Ting Lin and J. A. Sauls}
\affiliation{Department of Physics \& Astronomy \\
         Northwestern University, Evanston, IL 60208, USA}
\begin{abstract}
The low temperature properties of a wide range of many-fermion systems spanning metals, quantum gases and liquids to nuclear matter are well understood within the framework of Landau's theory of Fermi liquids. The low-energy physics of these systems is governed by interacting fermionic quasiparticles with momenta and energies near a Fermi surface in momentum space. Nonequilibrium properties are described by a kinetic equation for the distribution function for quasiparticles proposed by Landau.
Quasiparticle interactions with other quasiparticles, phonons or impurities lead to internal forces acting on a distribution of nonequilibrium quasiparticles, as well as collision processes that ultimately limit the transport of mass, heat, charge and magnetization, as well as limit the coherence times of quasiparticles.
For Fermi liquids that are close to a second order phase transition, e.g. Fermi liquids that undergo a superfluid transition, \emph{incipient} Cooper pairs - long-lived fluctuations of the ordered phase - provide a new channel for scattering quasiparticles, as well as corrections to internal forces acting on the distribution of nonequilibrium quasiparticles.
We develop the theory of quasiparticle transport for Fermi liquids in the vicinity of a BCS-type superfluid transition starting from Keldysh's field theory for non-equilibrium, strongly interacting fermions.
The leading corrections to Fermi liquid theory for non-equilibrium quasiparticle transport near a Cooper instability arise from the virtual emission and absorption of incipient Cooper pairs. 
Our theory is applicable to quasiparticle transport in superconductors, nuclear matter and the low temperature phases of liquid \He.
As an implementation of the theory we calculate the pairing fluctuation corrections to the attenuation of zero sound in liquid \He\ near the superfluid transition and demonstrate quantitative agreement with experimental results.
\end{abstract}
\maketitle
\section{Introduction}
The concept of quasiparticles introduced by Landau as the low-energy excitations of an interacting many-particle system~\cite{lan56} is a cornerstone of condensed matter physics, with applications spanning quantum fluids, conduction electrons in metals to nuclear matter.
Landau also introduced a kinetic equation for the distribution function of interacting quasiparticles in order to describe the nonequilibrium properties of Fermi systems~\cite{lan56,lan57}.
Landau's original formulation implicitly assumed that the ground state of a Fermi liquid was a state with unbroken symmetry defined by the Fermi sea and a quasiparticle vacuum, i.e. the absence of quasiparticles for states in the vicinity of the Fermi surface.
However, interactions between quasiparticles can lead to an instability of the Fermi liquid and a symmetry breaking phase transition to a new ordered ground state. In particular, liquid \He\ becomes a superfluid, and many metals become superconducting at low temperatures, driven by attractive interactions between quasiparticles.
It is thus interesting to study the corrections to Fermi liquid theory in the vicinity of a superfluid or superconducting phase transition.
When the temperature approaches the transition temperature from above, incipient Cooper pairs become long-lived providing a new excitations that can interact with and scatter quasiparticles. 
To formulate corrections to nonequilibrium dynamics resulting from quasiparticle-pair-fluctuation interactions we derive a kinetic equation starting from Keldysh's formulation of nonequilibrium field theory for an interacting Fermi system~\cite{kel65}. The theory is broadly applicable to intercting Fermi systems near a Cooper instability.

As an application of the theory presented here we consider the effects of pairing fluctuations on quasiparticle interactions and the attenuation of zero sound. The latter, predicted by Landau, is a Bosonic excitation of the Fermi surface, a coherent excitation of particles and holes that results in a propagating density and current oscilation of the Fermi liquid. The discovery of the zero sound, including measurements of the mode velocity and attenuation, provided powerful experimental confirmation that liquid \He\ is a Fermi liquid~\cite{abe66}.
At temperatures ranging from $T_c\approx 1-2.5\,\mbox{mK}$, depending on pressure, liquid \He\ undergoes a second-order phase transition to a condensate of spin-triplet, p-wave Cooper pairs, driven by the exchange of \emph{paramagnons}, i.e. long-lived ferromagnetic spin-fluctuations~\cite{lay71,lay74,and73,bri74,and75}.
The region of critical fluctuations, where order parameter fluctuations dominate, is extremely small for superfluid \He\ and nearly all superconductors. However, the region of Gaussian fluctuations, with a reduced phase space but divergent pair fluctuation propagaor, provides observable effects in a window of $(T-T_{\mathrm{c}})/T_{\mathrm{c}}\sim 10^{-2}$.
A classic signature of Gaussian fluctuations, well studied by many research groups, is the phenomenon of \emph{paraconductivity}, the decrease in resistivity as $T\rightarrow T_c^+$ attributed to enhanced conductivity from pair fluctuations~\cite{glo67,asl68a,mak68,mak68a,abr70,tho70,sko75,larkin05}.
By contrast the effects of Cooper pair fluctuations in liquid \He, in particular how these long-lived Bosonic excitations influence the nonequilibrium properties of \He\ above $T_{\mathrm{c}}$, have not been widely explored. 
Soon after the discovery of superfluidity in \He\ Emery proposed a phenomenological theory for scattering of quasiparticles off long-lived pair fluctuations and he calculated the corresponding corrections to the transport coefficients for normal liquid \He\ with the goal of identifying the orbital angular momentum of Cooper pairs in \He~\cite{eme76}.
Parpia et~al.~\cite{par78} reported deviations of the viscosity compared to the prediction of Fermi liquid theory, but only in a very narrow window of temperatures, $\sim 10^{-3}T_{\mathrm{c}}$. 
More substantial evidence of pairing fluctuations was provided by Paulson and Wheatley~\cite{pau78c} who observed enhanced attenuation of zero sound over a temperature window of order $\sim 10^{-1}T_{\mathrm{c}}$. This experiment motivated Samalam and Serene~\cite{sam78} to use Emery's collision integral to calculate the pair-fluctuation correction the zero sound attenuation, with a result that is in reasonable agreement with experiment.
However, as we demonstrate, the heuristic approach of Emery leads to an incorrect form for the quasiparticle-pair-fluctuation collision integral, in magnitude and more fundamentally. Emery's approach assumes that pair fluctuations with different quantum numbers are phase coherent and that their amplitudes add coherently. The microscopic theory for the collision integral that we derive contains no quantum interference terms between pair fluctuations with different quantum numbers because above $T_c$ these fluctuations are not phase coherent.
Indeed a key motivation for this work was to formulate a theory of quasiparticle-pair-fluctuation scattering for the nonequilibrium properties of a Fermi liquid near a superconducting/superfluid transition from a first-principles framework of quantum transport theory. 

The onset of superconductivity via the formation of long-lived Cooper pairs results from repeated scattering of pairs of quasiparticles with nearly zero total momentum and zero excitation energy, i.e. the \emph{ladder diagrams} in quantum field theory which generate the pair fluctuation propagator above $T_c$~\cite{AGD,larkin05}.
We show that the quasiparticle self-energy generated by the virtual exchange of pair fluctuations accounts for the leading-order correction to the Boltzmann-Landau kinetic equation for the quasiparticle distribution function.
Another factor motivating this work was the experimental observation of changes in the velocity of zero sound for temperatures near the superfluid transition in \He{}~\cite{lee96}. Pairing fluctuation corrections to the \emph{velocity} of zero sound are not accounted for in the formulation of Emery~\cite{eme76}, Samalam and Serene~\cite{sam78}, as shown by Pal and Bhattacharyya~\cite{pal79}. We discusss the theory for the quasiparticle-pair-fluctuation corrections to the zero sound velocity in a separate report~\cite{lin21a}. 
Lastly, we obtain corrections to the internal forces on quasiparticles arising from virtual emission and absorption of pair fluctuations which have not been studied before. Our hope is that these results will stimulate future research on fluctuations in liquid \He, as well as nonequilibrium quasiparticle transport properties in other Fermi liquids.

We derive a kinetic equation starting from Keldysh's formulation of nonequilibrium field theory in Sect.~\ref{sec-BLE}, including the leading self-energies that define Fermi liquid theory.
In Sect.~\ref{sec-fluctuation} we calculate the vertex function for incipient Cooper pairs within the Keldysh formalism, then develop the theory for p-wave pairing in liquid \He. The method is applicable to Cooper pairing for any orbital and spin representation. 
We then calculate the quasiparticle-pair-fluctuation self-energy in Sect.~\ref{sec-collision-integral} and show that this self-energy generates the leading corrections to Fermi-liquid theory, which become important as $T\rightarrow T_c^+$. Here we discuss the resulting quasiparticle-pair-fluctuation collision integral and compare with Emery's heuristic collision integral.
As an application of the theory, in Sect.~\ref{sec-zero-sound} we calculate the attenuation of zero sound in liquid \He\ due to scattering of pair fluctuations off the coherent particle-hole excitatation that is zero sound. 
We compare our theoretical results with the experimental measurements of zero sound by Paulson and Wheatley~\cite{pau78c}. The quantitative agreement between experiment and theory provides strong validation of our theory for the quasiparticle-pair-fluctuation self-energy.
Finally in Sect.~\ref{sec-param} we show that the same self-energy leads to corrections to the mean-field quasiparticle interaction energy, i.e. the Fermi-liquid self-energy. We derive results for the corrections to the Landau Fermi liquid parameters for liquid \He. 

\section{Fermi Liquid Theory}\label{sec-fermi-liquid}

We begin with a brief review to the microscopic formulation of Fermi liquid theory. Landau's original formulation of Fermi liquid theory implicitly assumed that the quasiparticle vacuum, i.e. the state no quasiparticles or holes, was the ground state, i.e. that even strong interactions would not lead to a symmetry breaking phase transition~\cite{lan56}.
However, many Fermi systems which exhibit the thermodynamic and transport properties of Fermi liquid theory undergo second-order phase transtions to superconducting or superfluid ground states. We thus need to develop Landau's Fermi liquid theory from a microscopic formulation in order to accommodate both the standard Landau interaction self-energies, as well as the interactions that drive the superconducting instability. This also allows us to calculate the corrections to Fermi liquid theory from long-lived fluctuations of the ordered phase in the near vicinity of the phase transition.

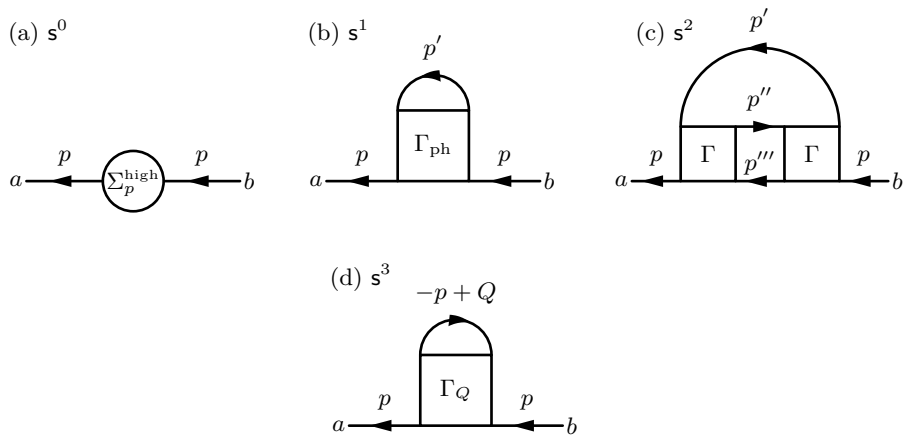
\begin{figure}[t]
\centering
\begin{fmffile}{fmf-Self_Energy_High}
  \begin{fmfgraph*}(100,45)
    \fmfpen{thin}
    \fmfset{arrow_len}{3mm}
    \fmftop{b1}
    \fmfv{label=(a) $\sml^0\hspace*{25mm}$}{b1}
    \fmfleft{l1,l2}
    \fmfright{r1,r2}
    \fmfv{label=$a$,l.d=0.5mm,l.a=180}{l1}
    \fmfv{label=$b$,l.d=0.5mm,l.a=0}{r1}
    \fmfv{decor.shape=circle,d.f=0,d.si=8mm,label=$\Sigma_{p}^{\mbox{\tiny high}}$,l.d=-1.5mm}{v1}
    \fmf{fermion,tension=.9,label=$p$,l.a=+90}{r1,v1}
    \fmf{fermion,tension=.9,label=$p$,l.a=-90}{v1,l1}
  \end{fmfgraph*}
\end{fmffile}
\begin{fmffile}{fmf-Self_Energy_Landau}
  \begin{fmfgraph*}(100,45)
    \fmfpen{thin}
    \fmfset{arrow_len}{3mm}
    \fmftop{b1}
    \fmfv{label=(b) $\sml^1\hspace*{25mm}$}{b1}
    \fmfleft{l1,l2}
    \fmfright{r1,r2}
    \fmfpolyn{empty,tension=.2,label=$\Gamma_{\mathrm{ph}}$}{G}{4}
    \fmf{fermion,tension=.9,label=$p$,l.a=+90}{r1,G1}
    \fmf{fermion,right,tension=0.1,label=$p'$}{G2,G3}
    \fmf{fermion,tension=.9,label=$p$,l.a=-90}{G4,l1}
    \fmfv{label=$a$,l.a=180,l.d=0.5mm}{l1}
    \fmfv{label=$b$,l.d=0.5mm,l.a=0}{r1}
  \end{fmfgraph*}
\end{fmffile}
\begin{fmffile}{fmf-Self_Energy_Binary}
  \begin{fmfgraph*}(120,45)
    \fmfpen{thin}
    \fmfset{arrow_len}{3mm}
    \fmftop{b1}
    \fmfv{label=(c) $\sml^2\hspace*{25mm}$}{b1}
    \fmfleft{l1,l2}
    \fmfright{r1,r2}
    \fmfv{label=$a$,l.d=0.5mm,l.a=180}{l1}
    \fmfv{label=$b$,l.d=0.5mm,l.a=0}{r1}
    \fmfpolyn{empty,tension=0.2,label=$\Gamma$}{G}{4}
    \fmfpolyn{empty,tension=0.2,label=$\Gamma$}{H}{4}
    \fmf{fermion,tension=0.9,label=$p$,l.a=90}{r1,H1}
    \fmf{fermion,tension=0.9,label=$p'''$,l.d=-4mm}{H4,G1}
    \fmf{fermion,tension=0.0,label=$p''$,l.a=90}{G2,H3}
    \fmf{fermion,right,tension=0.0,label=$p'$,l.a=90}{H2,G3}
    \fmf{fermion,tension=0.9,label=$p$,l.a=90}{G4,l1}
  \end{fmfgraph*}
\end{fmffile}
\\
\vspace{5em}
\begin{fmffile}{fmf-Self_Energy_Cooper}
  \begin{fmfgraph*}(100,45)
    \fmfpen{thin}
    \fmfset{arrow_len}{3mm}
    \fmftop{b1}
    \fmfv{label=(d) $\sml^3\hspace*{25mm}$}{b1}
    \fmfleft{l1,l2}
    \fmfright{r1,r2}
    \fmfv{label=$a$,l.d=0.5mm,l.a=180}{l1}
    \fmfv{label=$b$,l.d=0.5mm,l.a=0}{r1}
    \fmfpolyn{empty,tension=.2,label=$\Gamma_{Q}$}{G}{4}
    \fmf{fermion,tension=.9,label=$p$,l.a=90}{r1,G1}
    \fmf{fermion,left,tension=0.1,label=$-p+Q$}{G3,G2}
    \fmf{fermion,tension=.9,label=$p$,l.a=90}{G4,l1}
  \end{fmfgraph*}
\end{fmffile}
\caption{Self-energy diagrams and their order of magnitude in the small expansion 
         parameter, $\sml\in\{T_{\mathrm{c}}/E_\fm,1/k_\fm\xi_0\}$ of Fermi liquid theory.
(a) $\Sigma_{p}^{\mbox{\tiny high}}$ is $\cO(\sml^0)$ and gives the Fermi liquid mass 
    renormalization,
(b) is the Landau mean field self-energy of $\cO(\sml^1)$,
(c) is the self-energy from binary collision scattering and is $\cO(\sml^2)$, and
(d) is the self-energy derived scattering of quasiparticles by pair fluctuations and 
    is formally $\cO(\sml^3)$, but divergent for $T\rightarrow T_c^+$ and $Q\rightarrow 0$.
}
\label{fig-self-energies}
\end{figure}

Our analysis is based on the quasiclassical theory of interacting Fermi systems~\cite{ser83,rai94b,met98}. This formulation separates propagators and vertices into high- and low-energy parts relative to a cross-over energy scale $\Lambda$, measured relative to the Fermi energy. We can then re-sum the bare interactions and high-energy intermediate propagators into renormalized vertex functions that couple only low-energy propagators. An assumption of Fermi liquid theory is that there is no emergent low-energy scale associated with the renormalized vertices, and that their magnitudes are set by the high energy scale, e.g. the Fermi energy or interaction potential.
The parameter $\sml$ is used to represent the small ratios that govern the range of validity of Fermi liquid theory, e.g. $\Lambda/T_{\fm}, T_{\mathrm{c}}/E_{\fm}$, $1/k_{\fm}\xi_0$, etc.~\cite{ser83,rai94b}. In terms of $\sml$ the renormalized vertices are of $\cO(\sml^0)$. We can estimate the order of magnitude of self-energy diagrams built from these vertices and low-energy propagators in terms of $\sml$. Since the renormalized vertices are of $\cO(\sml^0)$, the estimate is determined by number of low-energy propagators and the phase space from internal summations confined to the low energy region~\cite{ser83,rai94b}.
For example, the two-point vertex, $\Sigma^{\mathrm{high}}$, shown in Figure~\refsub{fig-self-energies}{(a)} is the sum of all high-energy contributions to the self-energy and is classified as $\sml^0$. This self-energy generates the mass renormalization of bare fermions to the effective mass of fermionic quasiparticles.
Diagram~\refsub{fig-self-energies}{(b)} gives the Landau quasiparticle interaction energy with the Landau interaction represented by the vertex $\Gamma_{\mathrm{ph}}$, which contains one internal low-energy propagator and is estimated as
\begin{equation}
\Sigma \sim \int^{\Lambda} d\eps \int^{\Lambda} d\xi_{\V{p}} \frac{1}{\eps - \xi_{\V{p}}} 
\sim \Lambda \sim \cO(\sml^1),
\end{equation}
where the integral is restricted to a energy shell of width $\Lambda$ after the inclusion of all high-energy degrees of freedom into renormalized vertices. 
Similarly, diagram~\refsub{fig-self-energies}{(c)} describes collisions between quasiparticles and is of order $\sml^2$. This diagram is responsible for the finite lifetime of quasiparticles, or equivalently a collision rate of $1/\tau\propto T^2$. In the absence of disorder this coherence time for a quasiparticle excitation sets the baseline timescale for all transport properties in the Fermi liquid regime~\cite{bay78}. These self-energy contributions have been studied extensively in relation to Landau's Fermi liquid theory, c.f. Ref.~\cite{ser83}.
Self-energy terms involving higher numbers of low-energy intermediate are higher order in $\sml$ due to phase space restrictions. Thus, Fermi liquid theory can be defined by self-energy terms through order $\sml^2$, in which case we need only consider the 2-point and 4-point renormalized vertices.

The 4-point vertex represents the interaction between two quasiparticles and can be separated into two key scattering channels. One is the ``Landau channel'' in which the two incoming particles have an angle $\angle(\V{p}_1,\V{p}_2)\neq \pi$, while the other is the ``Cooper channel'', $\angle(\V{p}_1,\V{p}_2)=\pi$, where the two momenta are opposite~\cite{met98}.
The Cooper channel is singled out because repeated scattering of pairs of quasiparticles with zero total momentum  
leads to the Copper instability, and BCS condensation of Cooper pairs, for attractive interactions in the Cooper channel. 
The forward scattering limit of the Landau channel gives rise to the mean-field interaction energy of quasiparticles (diagram \refsub{fig-self-energies}{(b)}), while full renormalized vertex evaluated for all momenta on the Fermi surface generates the quasiparticle collision integral (diagram \refsub{fig-self-energies}{(c)}).
We use $\Gamma_{\mathrm{ph}}$ to denote the forward scattering limit of the 4-point vertex because there is a discontinuity as the momentum transfer $K=(k_0,\V{k})\to 0$, related to the order of limits, c.f. Refs.~\cite{AGD,ser83,met98}.

Our focus is on diagram~\refsub{fig-self-energies}{(d)}. The vertex $\Gamma_{Q}$ is obtained from the summation of ladder diagrams with two incoming particles having nearly opposite momenta near the Fermi surface with small total momentum $q\ll p_f$ (Eq.~\ref{eq-ladder_diagrams}). When the particle-particle irreducible vertex, $V$, is attractive $\Gamma_Q$ develops a pole and diverges at a critical temperature set by the bandwidth $\Lambda$, and the dimensionless pairing interaction, $g =-N(0)V$, $T_c=1.13\,\Lambda\,e^{-1/g}$. 
As shown below, diagram~\refsub{fig-self-energies}{(d)} is of order $\cO(\sml^3)$, but since since $\Gamma_{Q}$ is singular for $T\rightarrow T_c^+$ the self-energy from pair-fluctuations can lead to measureable corrections to the predictions of Fermi liquid theory in the vicinity of $T_{\mathrm{c}}$.

\section{Derivation of Boltzmann-Landau Equation}\label{sec-BLE}

\begin{figure}[tbp]
\centerline{\includegraphics[width=1.0\textwidth]{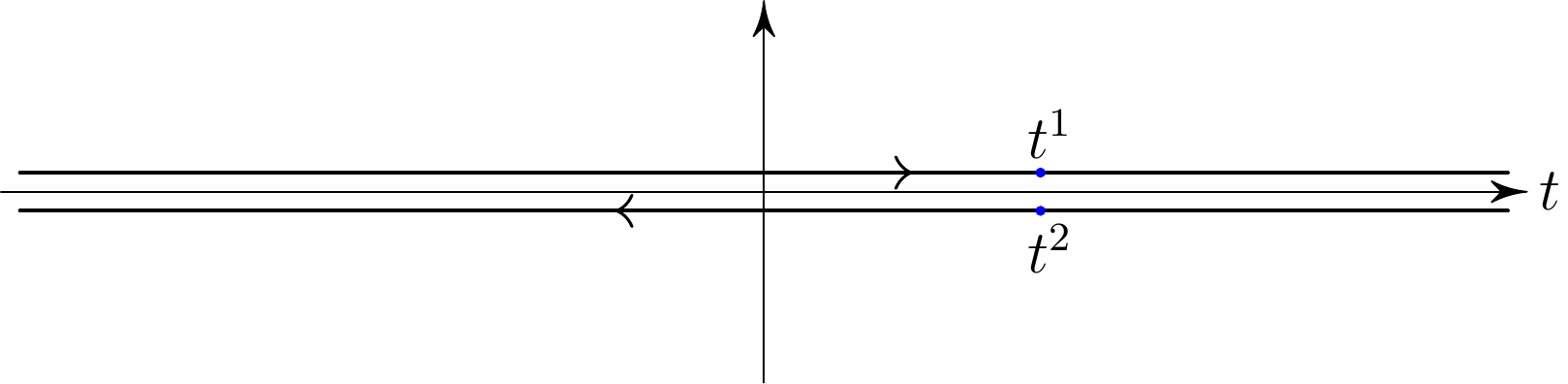}}
\caption[]{The Keldysh contour. Time coordinates shown on the upper (lower) branch with Keldysh index $1$ ($2$).}
\label{keldysh}
\end{figure}

We formulate the nonequilibrium transport equation for interacting fermions using Keldysh method~\cite{kel65}. The kinetic equation of Fermi liquids can be derived using nonequilibrium Green functions~\cite{ser83,rai94b,kit10}. Consider the Keldysh contour $\mathcal{C}$ where the time variable goes from $-\infty$ to $+\infty$ and comes back to $-\infty$, as shown in Figure~\ref{keldysh}.
%
%
We define the contour-ordered Green function 
\begin{equation}
G(1^a, 2^b) \equiv -i \langle \mathsf{T}_{\mathcal{C}} \psi(1^a) \psi^{\dagger}(2^b) \rangle ,
\end{equation}
where the expression $\psi(i)$ represents $\psi(x_i,t_i)$ with $x$ includes the spatial and spin variables, and the superscripts $a,b$ represent the forward and backward branches of the Keldysh contour, indexed by $1$ and $2$ respectively. Dyson's equation of the contour-ordered Green function~\cite{kit10} is
\begin{equation}
\Big(i\frac{\partial}{\partial t_1^a} - K_1 \Big)G(1^a, 2^b)
- \int_{3^c} \Sigma(1^a, 3^c) G(3^c, 2^b) d3^c 
= \delta^\mathrm{C}(1^a, 2^b),
\end{equation}
where $K_1$ represents the single-particle Hamiltonian, $K_1 \equiv \nabla_1^2/2m - \mu$, operating on the first variable $1$, and $\Sigma$ is the self-energy due to interactions. The contour delta function is defined as
\begin{equation}
\delta^\mathrm{C}(t^a, t'^b) =
\begin{cases}
\delta(t-t'),& a=b=1 \\
-\delta(t-t'),& a=b=2 \\
0,& a\neq b
\end{cases}.
\end{equation}
The minus sign in this definition is due to the Heaviside step function in the anti-time-ordered Green function $G^{\tilde{\mathrm{T}}}$. Note that the notation $\int_{3^c}$ represents the summation over all dummy variables, including a time integral along the Keldysh contour.

The relation between the contour-ordered Green function and the conventional Green functions is given by the matrix $\check{G}^{ab}(1,2)\equiv G(1^a, 2^b)$. Using the definition of contour-ordering, it is easy to establish that
\begin{equation}
\check{G} =
\begin{pmatrix}
G^{\mathrm{T}} & G^{<} \\
G^{>} & G^{\tilde{\mathrm{T}}}
\end{pmatrix}.
\end{equation}
The same definition is used for the self-energy $\Sigma(1^a, 3^c)$. Recall that the contour time integral includes a forward and a backward time integrals. Thus we have Dyson equations for each $\check{G}^{ab}$:
\begin{align}
&\Big(i\frac{\partial}{\partial t_1} - K_1\Big) G^{11} 
-\int \big[\Sigma^{11}G^{11} - \Sigma^{12}G^{21}\big] = \delta(1, 2), 
\\
\label{left-dyson}
&\Big(i\frac{\partial}{\partial t_1} - K_1\Big) G^{12}
- \int \big[\Sigma^{11}G^{12} - \Sigma^{12}G^{22}\big] = 0,
\\
&\Big(i\frac{\partial}{\partial t_1} - K_1\Big) G^{21}
- \int \big[\Sigma^{21}G^{11} - \Sigma^{22}G^{21}\big] = 0,
\\
&\Big(i\frac{\partial}{\partial t_1} - K_1\Big) G^{22}
- \int \big[\Sigma^{21}G^{12} - \Sigma^{22}G^{22}\big] = -\delta(1, 2),
\end{align}
where the symbol $\int$ represents convolutions for all intermediate variables, including spins, with the standard time integral from $-\infty$ to $\infty$. In the above Dyson equations, the differential operators are applied to the first variable of the Green function, and thus they are called left Dyson equations. We can rewrite the Dyson equations by applying operators to the second variables, and obtain the right Dyson equations. Consider the lesser Green function $G^{12}$. The corresponding right Dyson equation is
\begin{equation}
\label{right-dyson}
\Big(-i\frac{\partial}{\partial t_2} - K_2\Big) G^{12}
- \int \big[G^{11}\Sigma^{12} - G^{12}\Sigma^{22}\big] = 0.
\end{equation}
Subtracting the left Dyson equation~\eqref{left-dyson} from the right Dyson equation~\eqref{right-dyson}, we obtain
\begin{equation}
\label{LRsubt}
\hspace*{-1em}
\Big(-i\big(\frac{\partial}{\partial t_1} + \frac{\partial}{\partial t_2}\big)
+ \frac{1}{2m}\big( \nabla^2_2 - \nabla_1^2  \big)
\Big) G^{12} \\
- \int 
\Big[
G^{11}\Sigma^{12} - G^{12}\Sigma^{22}
-\Sigma^{11}G^{12} + \Sigma^{12}G^{22}
\Big] = 0.
\end{equation}

When the system is homogeneous in both space and time, all Green functions depend only on the differences $\mathbf{r_1-r_2}$ and $t_1 - t_2$. It is then natural to describe the system in the Fourier space. If an external perturbation is long-wavelength and low-frequency compared to the Fermi wavelength and the Fermi energy, we can simplify the equations using the separation of scales. For convenience, consider the mixed coordinates $\mathbf{R}\equiv(\mathbf{r_1+r_2})/2$, $\mathbf{r}\equiv\mathbf{r_1-r_2}$, and $T\equiv(t_1+t_2)/2$, $t\equiv t_1-t_2$. More compactly, we write $X\equiv(T,\mathbf{R})$, $x\equiv(t,\mathbf{r})$. The Wigner transformation is defined as 
\begin{equation}
G(X, p) = \int dx e^{-ipx}G(X+\frac{x}{2}, X-\frac{x}{2}),
\end{equation}
where $p\equiv(\epsilon, \mathbf{p})$ is the corresponding Fourier variable, and $px\equiv-\epsilon t + \mathbf{p}\cdot\mathbf{r}$ is the inner product between $p$ and $x$. This transformation keeps the center-of-mass coordinates unchanged, and Fourier transforms the relative coordinates. When the system is homogeneous in space and time, i.e.~independent of $X$, this transformation is reduced to the usual Fourier transformation.
Using the Wigner representation of the Green functions, the first term in~\eqref{LRsubt} can be written as
\begin{equation}
\label{drift}
\Big(-i\big(\frac{\partial}{\partial t_1} + \frac{\partial}{\partial t_2}\big)
+ \frac{1}{2m}\big( \nabla^2_2 - \nabla_1^2  \big)
\Big) G^{12} 
\implies
\big(
\partial_T
+ \frac{\mathbf{p}}{m}\cdot\nabla_\mathbf{R}
\big)[-iG^{12}].
\end{equation}
The second term contains convolutions, whose spatial and temporal parts in the mixed coordinates are given by Moyal products\cite{moy49},
\begin{equation}
(A\otimes B)(X, p) =
\exp\Big[\frac{i}{2}\big(\partial^A_X\partial^B_p - \partial^A_p \partial^B_X \big)\Big]
A(X,p)B(X,p),
\end{equation}
where
\begin{equation}
\partial^A_X \partial^B_p \equiv
-\partial^A_T \partial^B_\epsilon + \partial^A_\mathbf{R} \partial^B_\mathbf{p},
\end{equation}
and the upper indices $A$ and $B$ indicate the function the derivative operator applies to. Here we focus on systems in the absence of a magnetic field, so the spin structure is trivial and omitted hereafter. Since we are interested in the long-wavelength behavior, we also keep only the zero- and first-order terms in the expansion of Moyal product,
\begin{equation}\label{grad-expansion}
\begin{split}
(A\otimes B) &\approx AB + \frac{i}{2}
\big[ \partial_X A \partial_p B - \partial_p A \partial_X B \big] \\
&\equiv AB + \frac{i}{2}\{A, B\}_\mathrm{PB}.
\end{split}
\end{equation}
Using~\eqref{drift},~\eqref{grad-expansion} and the relations $G^{11}+G^{22}=G^{12}+G^{21}$, $\Sigma^{11}+\Sigma^{22}=\Sigma^{12}+\Sigma^{21}$, the equation~\eqref{LRsubt} can be written as
\begin{multline}\label{QBE}
\big(
\partial_T
+ \frac{\mathbf{p}}{m}\cdot\nabla_\mathbf{R}
\big)[-iG^{12}] 
- \big[G^{21}\Sigma^{12} - G^{12}\Sigma^{21}\big] 
-\frac{i}{2}
\{G^{11}-G^{22}, \Sigma^{12}\}_\mathrm{PB} 
+ \frac{i}{2}
\{\Sigma^{11} - \Sigma^{22}, G^{12}\}_\mathrm{PB}
=0.
\end{multline}

To make a connection with the quasiclassical Boltzmann-Landau kinetic equation, we consider the distribution function defined by
\begin{equation}
n(X, p)\equiv \frac{-iG^{12}(X,p)}{A(X, p)},
\end{equation}
where $A(X, p)$ is the local spectral function. In equilibrium, it can be shown\cite{kadanoff76} that the distribution is indeed the Fermi-Dirac distribution $n(X, p) = 1/(e^{\beta\epsilon}+1)$, where $\beta=1/T$ is the inverse temperature of the bath. Thus, the function $n$ defined above can be interpreted as a distribution function. We assume the spectral function can be approximated by a quasiparticle peak when the energy is much lower than the Fermi energy. More precisely, the lesser Green function can be approximated by $-iG^{12}\approx2\pi\delta(\eps - \eps_\V{p})n$. Thus the integration over $\epsilon$ in the low-energy regime gives the distribution function,
\begin{equation}
\int_{-\epsilon_\mathrm{c}}^{\epsilon_\mathrm{c}}\frac{d\epsilon}{2\pi} (-iG^{12})
\approx n|_{\epsilon=\epsilon_{\mathbf{p}}}
\equiv n_{\V{p}}(\V{R},T).
\end{equation}
By integrating equation~\eqref{QBE} with respect to energy $\eps$, we can obtain
\begin{multline}
\label{BLE}
(\partial_t + \frac{\V{p}}{m}\cdot \nabla_\V{R})n_{\V{p}}
- \nabla_\V{R}\mathfrak{Re}\Sigma^{11}\big|_{\eps=\eps_\V{p}}\cdot\nabla_\V{p}n_{\V{p}} 
+ \nabla_\V{p}\mathfrak{Re}\Sigma^{11}\big|_{\eps=\eps_\V{p}}\cdot\nabla_\V{R}n_{\V{p}}
- \partial_\eps \mathfrak{Re}\Sigma^{11}\big|_{\eps=\eps_\V{p}} \partial_t n_{\V{p}}  \\
=-i\Sigma^{12}\big|_{\eps=\eps_\V{p}}(1-n_{\V{p}}) - i\Sigma^{21}\big|_{\eps=\eps_\V{p}}n_{\V{p}},
\end{multline}
where we have changed the time variable $T\to t$, and the substitution $\eps\to \eps_{\V{p}}$ is performed after the differentiation. Using quasiclassical methods~\cite{ser83,rai94b}, the leading order contribution to the self-energy, $\mathfrak{Re}\Sigma^{11}$, is a mean-field energy, and the $\Sigma^{12}$ and $\Sigma^{21}$ are given by the second order diagram~\refsub{fig-self-energies}{(c)}. These two terms give rise to Landau's quasiclassical kinetic equation, which is the cornerstone of nonequilibrium properties of Fermi liquids. However, when the system is close to the transition temperature the vertex $\Gamma_Q$ in the Cooper channel is singular leading to singular corrections to Boltzmann-Landau kinetic equation.

\section{Cooper Pair Fluctuations}\label{sec-fluctuation}

In this section we calculate the vertex $\Gamma_Q$ by summing ladder diagrams of pairs of quasiparticles scattering with almost opposite momenta. Since this is the only relevant vertex discussed hereafter we drop the $Q$ in the subscript. The Cooper instability is due to an attractive interaction between quasiparticles with opposite momenta on the Fermi surface. 
As a specific example we consider a spin-triplet, p-wave pairing interaction which is applicable to the pairing instability and superfluidity in liquid \He~\cite{vollhardt90}. In this case the pairing interaction is of the form, 
\begin{equation}
V_{\alpha\beta, \gamma\delta}(p,p')=
3V\hat{\V{p}}\cdot\hat{\V{p}}' \times
\frac{1}{2}\V{g}_{\alpha\beta}\cdot\V{g}^\dagger_{\gamma\delta}
\end{equation}
with the coupling constant $V<0$ and $\V{g} \equiv i\vsigma\sigma_y$ is the triplet of spin-symmetric matrices for spin-triplet pairing. The vertex function for the Cooper instability is given by the following Bethe-Salpeter equation
%
\vspace{2em}
\begin{equation}\label{eq-ladder_diagrams}
\hspace{-1em}
\begin{fmffile}{ladder}
  \begin{gathered}
  \begin{fmfgraph*}(60,40)
    \fmfpen{thin}
    \fmfset{arrow_len}{3mm}
    \fmfleft{l1,l2}
    \fmfright{r1,r2}
    \fmfpolyn{empty,tension=.5,label=$\Gamma$}{G}{4}
    \fmf{fermion,tension=.5}{r1,G1}
    \fmf{fermion,tension=.5}{r2,G2}
    \fmf{fermion,tension=.5}{G3,l2}
    \fmf{fermion,tension=.5}{G4,l1}
    \fmfv{label=$p$}{l1}
    \fmflabel{$Q-p$}{l2}
    \fmflabel{$Q-p'$}{r2}
    \fmflabel{$p'$}{r1}
  \end{fmfgraph*}
  \end{gathered}
  \hspace{2em} = \hspace{2em}
  \begin{gathered}
  \begin{fmfgraph*}(60,40)
    \fmfpen{thin}
    \fmfset{arrow_len}{3mm}
    \fmfleft{l1,l2}
    \fmfright{r1,r2}
    \fmfpolyn{empty,tension=.5,label=$V$}{V}{4}
    \fmf{fermion,tension=.5}{r1,V1}
    \fmf{fermion,tension=.5}{r2,V2}
    \fmf{fermion,tension=.5}{V3,l2}
    \fmf{fermion,tension=.5}{V4,l1}
    \fmflabel{$p$}{l1}
    \fmflabel{$Q-p$}{l2}
    \fmflabel{$Q-p'$}{r2}
    \fmflabel{$p'$}{r1}
  \end{fmfgraph*}
  \end{gathered}
  \hspace{2em} + \hspace{2em}
  \begin{gathered}
    \begin{fmfgraph*}(120,40)
    \fmfpen{thin}
    \fmfset{arrow_len}{3mm}
    \fmfleft{l1,l2}
    \fmfright{r1,r2}
    \fmfpolyn{empty,tension=.5,label=$\Gamma$}{G}{4}
    \fmfpolyn{empty,tension=.5,label=$V$}{V}{4}
    \fmf{fermion,tension=.8}{r1,G1}
    \fmf{fermion,tension=.8}{r2,G2}
    \fmf{fermion,right=0.5,tension=.4,label=$Q-p''$}{G3,V2}
    \fmf{fermion,left=0.5,tension=.4,label=$p''$}{G4,V1}
    \fmf{fermion,tension=.8}{V3,l2}
    \fmf{fermion,tension=.8}{V4,l1}
    \fmflabel{$p$}{l1}
    \fmflabel{$Q-p$}{l2}
    \fmflabel{$Q-p'$}{r2}
    \fmflabel{$p'$}{r1}
    \end{fmfgraph*}
  \end{gathered}.
\end{fmffile}
\vspace{2em}
\end{equation}
%
Using the Feynman rules applicable to Keldysh propagators we obtain\cite{kit10},
\begin{multline}
\Gamma_{\alpha\beta, \gamma\delta}^{ab}(p,p';Q) =
iV_{\alpha\beta, \gamma\delta}(p,p')\check{\tau}_3^{ab} \\
+ \sum iV_{\alpha\beta, \alpha'\beta'}(p,p'')\check{\tau}_3^{ac}
G^{cd}(p'')G^{cd}(Q-p'')
\Gamma_{\alpha'\beta',\gamma\delta}^{db}(p'', p';Q),
\end{multline}
where the superscripts $a,b,c,d$ are Keldysh indices, and the summation includes all repeated dummy indices. The symbol $p$ denotes fermion four-momentum, while the momentum $Q=(\V{q},\omega)$ is the total four-momentum of the Cooper pair. The resulting equation can be simplified isolating the spin-dependence for pure spin-exchange interactions,
\begin{equation}
\Gamma_{\alpha\beta, \gamma\delta}^{ab}(p,p';Q) = \Gamma^{ab}(p,p';Q)
\times 
\frac{1}{2}\V{g}_{\alpha\beta}\cdot\V{g}^\dagger_{\gamma\delta},
\end{equation}
which separates spin and orbital degrees of freedom. The Bethe-Salpeter equation reduces to
\begin{equation}\label{BS1}
\Big[\frac{1}{3iV}\check{\tau}^{ac}_3 \Gamma^{cb}- 
\sum \hat{\V{p}}\cdot\hat{\V{p}}''
G^{ac}(p'')G^{ac}(Q-p'') 
\Gamma^{cb}(p'',p';Q)\Big]
=\hat{\V{p}}\cdot\hat{\V{p}}'\delta^{ab}.
\end{equation}
Orbital rotation symmetry allows us to further separate the vertex function into two components of the uniaxial 
orbital structure determined by $\V{q}$,
\begin{equation}
\Gamma^{ab} = \sum_{\lambda=\parallel,\perp}\hat{p}_i [ \Gamma_{\lambda}^{ab}P_{ij}^{\lambda}] \hat{p}'_j
\end{equation}
with the two components determined the uniaxial tensors, $P_{ij}^{\parallel}=\hat{q}_i\hat{q}_j$ and $P_{ij}^{\perp}=\delta_{ij}-\hat{q}_i\hat{q}_j$. The vertex function $\Gamma$ depends on the direction of quasiparticle momenta, but is insensitive to the quasiparticle energies. 
Integrating out the frequency $\epsilon''$, the summation over $p''\equiv(\epsilon'',\V{p}'')$ can then be expressed in
terms of the expressions,
\begin{align}
\label{GG11}
(GG)^{11}&= i\int_{\mathbf{p}} 
\frac{n_{\mathbf{p}}+n_{\mathbf{q-p}}-1}{\omega - \eps_\mathbf{p}-\eps_\mathbf{q-p}+i0}
\hat{p}_i \hat{p}_k
+(GG)^{12},
\\
\label{GG12}
(GG)^{12} &= -2\pi \int_\mathbf{p}\delta(\omega - \eps_\mathbf{p} - \eps_\mathbf{q-p})
n_{\mathbf{p}}n_{\V{q}-\V{p}}
\hat{p}_i \hat{p}_k,
\\
\label{GG21}
(GG)^{21} &= -2\pi \int_\mathbf{p} \delta(\omega - \eps_\mathbf{p} - \eps_\mathbf{q-p})
\bar{n}_{\mathbf{p}}\bar{n}_{\V{q}-\V{p}}
\hat{p}_i \hat{p}_k,
\\
\label{GG22}
(GG)^{22} &= -i\int_\mathbf{p} 
\frac{n_{\mathbf{p}}+n_{\V{q}-\V{p}}-1}{\omega - \eps_\mathbf{p}-\eps_\mathbf{q-p}+i0}
\hat{p}_i \hat{p}_k
+(GG)^{21},
\end{align}
where we introduced 
\begin{equation}
(GG)^{ac}_{ik}\equiv\sum_{p''} \hat{p}''_{i} G^{ac}(p'')G^{ac}(Q-p'')\hat{p}''_k
\,.
\end{equation}
We also changed the dummy variable $\V{p}''\to\V{p}$ in the integrals for simplicity, and we defined $\bar{n}_{\V{p}} \equiv 1 - n_{\V{p}}$. Note that the distribution function $n_{\V{p}} \equiv n(\eps_{\V{p}},\V{p})$ is in general the nonequilibrium quasiparticle distribution.

Furthermore, we can decompose $(GG)$ into $\parallel$ and $\perp$ components: $(GG)^{ac}_{ik}=\sum_{\lambda=\parallel,\perp} (GG)^{ac}_\lambda P_{ik}^{\lambda}$. Thus, Eq.~\eqref{BS1} reduces to 
\begin{equation}
\Big[\frac{1}{3iV}\check{\tau}_3^{ac} -
(GG)^{ac}_\lambda \Big]\Gamma^{cb}_\lambda = 1\delta^{ab}
\end{equation}
for $\lambda = \parallel, \perp$. Defining
\begin{equation}\label{LGL}
L_{ik} \equiv \frac{1}{3iV} 
- i\int_\mathbf{p} 
\frac{n_{\mathbf{p}}+n_{\V{q}-\V{p}}-1}{\omega - \eps_\mathbf{p}-\eps_\mathbf{q-p}+i0}
\hat{p}_i \hat{p}_k \,,
\end{equation}
we can express $L_{ij}=\sum_{\lambda} L^{\mathrm{GL}}_{\lambda} P^{\lambda}_{ij}$. For an equilibrium distribution of quasiparticles undergoing scattering in the Cooper channel we obtain
\begin{equation}\label{singular}
L^\mathrm{GL}_\lambda = i\frac{N(0)}{3}\Big(\vartheta + \xi_\lambda^2 q^2 - i\frac{\pi}{8}\frac{\omega}{T}\Big),
\end{equation}
with the reduced temperature $\vartheta\equiv T/T_{\mathrm{c}}-1$ and the coherence length scales, $\xi_\parallel^2 = \frac{9}{5}\xi_0^2$ and $\xi_\perp^2 = \frac{3}{5}\xi_0^2$, where $\xi_0^2 = \frac{7\zeta(3)}{48\pi^2}\frac{v_\mathrm{F}^2}{T_\mathrm{c}^2}$. The latter corresponds to the spin-singlet, s-wave coherence length entering the fluctuation propagator. This expression for $L^{\mathrm{GL}}_{\lambda}$ is the same as the result we obtain using the Matsubara method~\cite{larkin05}. The above matrix equation can be written as $L_{\lambda}\Gamma_{\lambda} = I$ with
\begin{equation}
L_\lambda \equiv
\begin{bmatrix}
L_\lambda^\mathrm{GL} - (GG)^{12}_\lambda & -(GG)^{12}_\lambda \\
-(GG)^{21}_\lambda & -L_\lambda^\mathrm{GL} - (GG)^{21}_\lambda
\end{bmatrix}.
\end{equation}
The inverse gives the Cooper channel vertex function
\begin{equation}
\begin{split}
\hspace*{-1em}
\Gamma_\lambda =& L_\lambda^{-1} \\
=&\frac{1}{|L^{\mathrm{GL}}_\lambda|^{2} }
\begin{bmatrix}
-L^\mathrm{GL}_\lambda - (GG)^{21}_\lambda & (GG)^{12}_\lambda \\
(GG)^{21}_\lambda & L_\lambda^\mathrm{GL} - (GG)^{12}_\lambda
\end{bmatrix},
\end{split}
\end{equation}
where we used the identity
\begin{equation}
(GG)^{12}_{\lambda} - (GG)^{21}_{\lambda}  = L^{\mathrm{GL}}_{\lambda} + (L^{\mathrm{GL}}_{\lambda})^{*},
\end{equation}
which can be seen from~\eqref{GG11}--\eqref{GG22} and~\eqref{LGL}. The factor $1/|L^{\mathrm{GL}}_{\lambda}|^2$ is 
singular at $q=0$, $\omega=0$ and $\vartheta\rightarrow 0^+$, signifying the onset of the superfluid transition.

\section{Collision Integral}\label{sec-collision-integral}

With this result for the pair propagator we can evaluate the quasiparticle-pair-fluctuation self-energy. Here we consider the collision integral derived from the scattering of quasiparticles by long-lived Cooper pair fluctuations.
We separate the non-collisional self-energies to the left side of the transport equation, and in this case retain only the 
leading order molecular field self-energy of order $\sml^1$ from diagram \refsub{fig-self-energies}{(b)}. Thus, the left-hand side of~\eqref{BLE} reduces to that of the standard Boltzmann-Landau kinetic equation
\begin{equation}
\mathrm{LHS} = \partial_t n_{\V{p}} + \frac{\V{p}}{m^{*}}\cdot\nabla_{\V{R}} n_{\V{p}} 
             - \nabla_{\V{R}}\epsilon_{\V{p}}\cdot\nabla_{\V{p}} n_{\V{p}}
\,,
\end{equation}
where $m^{*}$ is the quasiparticle effective mass, and
\begin{equation}\label{quasiparticle-energy}
\eps_{\V{p}} = \eps_{\V{p}}^0 + 
\frac{1}{V} \sum_{\V{p}'\sigma'} f_{\V{p}\sigma, \V{p}'\sigma'} \delta n_{\V{p}'\sigma'}
\,,
\end{equation}
where the sum involving $f_{\V{p}\sigma,\V{p}'\sigma'}$ is the mean-field quasiparticle interaction energy. As for the right-hand side of the kinetic equation, the terms $\Sigma^{21}$ and $\Sigma^{12}$ connect outgoing and incoming lines of the vertex $\Gamma$ obtained in the previous section, as shown in Figure~\refsub{fig-self-energies}{(d)}, and generate the ''scattering out'' and ''scattering in'' contributions to the quasiparticle collision integral, respectively.
%
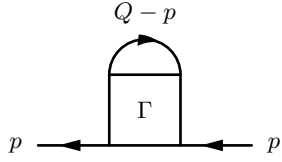
\begin{figure}[tbp!]
\centering
\begin{fmffile}{selfenergy}
  \begin{fmfgraph*}(100,70)
    \fmfpen{thin}
    \fmfset{arrow_len}{3mm}
    \fmfleft{l1,l2}
    \fmfright{r1,r2}
    \fmfpolyn{empty,tension=.2,label=$\Gamma$}{G}{4}
    \fmf{fermion,tension=.9}{r1,G1}
    \fmf{fermion,left,tension=0.1,label=$Q-p$}{G3,G2}
    \fmf{fermion,tension=.9}{G4,l1}
    \fmfv{label=$p$,l.a=180}{l1}
    \fmfv{label=$p$,l.a=0}{r1}
  \end{fmfgraph*}
\end{fmffile}
\caption{The quasiparticle-pair-fluctuation self-energy.}
\label{selfenergy}
\end{figure}
%
Thus, we have
\begin{equation}\label{sigma12}
\begin{split}
\Sigma^{12}(p)=\sum_Q \Gamma^{12}_{\alpha\beta,\gamma\delta}(p,p;Q)G^{21}_{\beta\delta}(Q-p)
\end{split}
\,,
\end{equation}
\begin{equation}\label{sigma21}
\begin{split}
\Sigma^{21}(p)=\sum_Q \Gamma^{21}_{\alpha\beta,\gamma\delta}(p,p;Q)G^{12}_{\beta\delta}(Q-p)
\,.
\end{split}
\end{equation}
Note that the vertex and the Green functions are functionals of the nonequilibrium distribution function. From Sect.~\ref{sec-fluctuation}, the $\Gamma^{12}$ sector of the vertex without spin dependence is given by
\begin{equation}\label{eq-Gamma12}
\Gamma^{12} = \sum_{\lambda=\parallel,\perp}
\frac{1}{|L^{\mathrm{GL}}_{\lambda}|^2}
\hat{p}_i (GG)_{\lambda}^{12}P^\lambda_{ij} \hat{p}'_j
\end{equation}
with $(GG)_\parallel=\sum_{i,j}(GG)_{ij}P^\parallel_{ji}$ and $(GG)_\perp=\frac{1}{2}\sum_{i,j}(GG)_{ij}P^\perp_{ji}$. Using Eq.~\eqref{GG12} for $(GG)^{12}$ we obtain
\begin{equation}
(GG)_{\lambda}^{12} =
-2\pi \int_\mathbf{p''}\delta(\omega - \eps_\mathbf{p''} - \eps_\mathbf{q-p''})
n(\eps_\mathbf{p''})n(\omega - \eps_\mathbf{p''}) 
\begin{cases}
(\hat{\V{p}}''\cdot\hat{\V{q}})^2\,, & \lambda=\parallel 
\\
\frac{1}{2}(1-(\hat{\V{p}}''\cdot\hat{\V{q}})^2)\,, & \lambda=\perp
\end{cases}\,.
\end{equation}

The natural quantization axis for the pair fluctuations is $\hat{\V{q}}||\hat{\V{z}}$, in which case we express the fluctuation propagators in terms of the spherical harmonics,
\begin{align}
\hat{p}_i P^{\parallel}_{ij} \hat{p}'_{j}
&= (\hat{\V{p}}\cdot\hat{\V{q}})(\hat{\V{p}}'\cdot\hat{\V{q}})
= \frac{4\pi}{3} Y_{10}(\hat{\V{p}})Y_{10}^*(\hat{\V{p}}')\,,
\\
\hat{p}_i P^{\perp}_{ij} \hat{p}_j'
&=(\hat{\V{p}}\cdot\hat{\V{p}}')-(\hat{\V{p}}\cdot\hat{\V{q}})(\hat{\V{p}}'\cdot\hat{\V{q}})
= \frac{4\pi}{3}\Big[Y_{11}(\hat{\V{p}})Y_{11}^*(\hat{\V{p}}')+Y_{1,-1}(\hat{\V{p}})Y_{1,-1}^*(\hat{\V{p}}')\Big]
\,.
\end{align}
Thus, we have 
\begin{multline}
\hat{p}_i (GG)_{\lambda}^{12}P^\lambda_{ij}\hat{p}'_j
=-2\pi\int_\mathbf{p''}\delta(\omega-\eps_\mathbf{p''}-\eps_\mathbf{q-p''})
n(\eps_\mathbf{p''})n(\omega - \eps_\mathbf{p''})
\\
\times 
\begin{cases}
\big(\frac{4\pi}{3}\big)^2 Y_{10}(\hat{\V p}'')Y_{10}^*(\hat{\V p}'')
Y_{10}(\hat{\V p})Y_{10}^*(\hat{\V p}')\,, & \lambda = \parallel \\
\frac{1}{2}\big(\frac{4\pi}{3}\big)^2 [Y_{11}Y_{11}^* + Y_{1,-1}Y_{1,-1}^*]
[Y_{11}Y_{11}^* + Y_{1,-1}Y_{1,-1}^*]\,, & \lambda = \perp 
\end{cases}\,,
\end{multline}
where the spherical harmonics for $\lambda=\perp$ are evaluated with same momentum variables as the components with $\lambda=\parallel$. The denominator in Eq.~\eqref{eq-Gamma12}, $|L^{\mathrm{GL}}_{\lambda}|^2$, can be evaluated in equilibrium and is given by Eq.~\eqref{singular}. 
For the self-energy we require $\V{p}=\V{p}'$. Then using 
$Y_{11}(\hat{\V{p}})Y_{11}^*(\hat{\V{p}})=Y_{1,-1}(\hat{\V{p}})Y_{1,-1}^*(\hat{\V{p}})$ 
we obtain,
\begin{multline}
\hat{p}_i (GG)_{\lambda}^{12}P^\lambda_{ij} \hat{p}'_j
=-2\pi \int _\mathbf{p''}\delta(\omega-\eps_\mathbf{p''}-\eps_\mathbf{q-p''})
n(\eps_\mathbf{p''})n(\omega - \eps_\mathbf{p''}) \\
\times 
\begin{cases}
\big(\frac{4\pi}{3}\big)^2 Y_{10}(\hat{\V p}'')Y_{10}^*(\hat{\V p}'')
Y_{10}(\hat{\V p})Y_{10}^*(\hat{\V p}), & \lambda = \parallel \\
2\big(\frac{4\pi}{3}\big)^2 Y_{11}(\hat{\V p}'')Y_{11}^*(\hat{\V p}'')
Y_{11}(\hat{\V p})Y_{11}^*(\hat{\V p}), & \lambda = \perp
\end{cases}.
\end{multline}
Note all three orbital components of the Cooper pair fluctuations contribute with equal weight. 
The factor of $2$ in the $\lambda=\perp$ component originates from the degenerate $m=\pm 1$ modes. 
Defining
\begin{equation}
L^{\mathrm{GL}}_m 
= 
\begin{cases}
L^{\mathrm{GL}}_\parallel, & m=0
\\
L^{\mathrm{GL}}_\perp,     & m=\pm 1 
\end{cases},
\end{equation}
we obtain
\begin{align}
&\Gamma^{12}(p,p;Q) = \sum_{\lambda=\parallel,\perp}
\frac{1}{|L^{\mathrm{GL}}_{\lambda}|^2}
\hat{p}_i (GG)_{\lambda}^{12}P^\lambda_{ij} \hat{p}_j \notag \\
&= - 2\pi \Big(\frac{4\pi}{3}\Big)^2 \int _\mathbf{p''}\delta(\omega - \eps_\mathbf{p''} - \eps_\mathbf{q-p''})
f(\eps_\mathbf{p''})f(\omega - \eps_\mathbf{p''})
\sum_m\bigg| 
\frac{1}{L^{\mathrm{GL}}_{m}} Y_{1m}(\hat{\V p})Y_{1m}^*(\hat{\V p}'')
\bigg|^2.
\end{align} 
The full Cooper pair fluctuation vertex, including the spin part, is then 
\begin{equation}\label{full-vertex}
\Gamma_{\alpha\beta, \gamma\delta}^{ab}(p,p';Q) = \Gamma^{ab}(p,p';Q)
\times 
\frac{1}{2}\V{g}_{\alpha\beta}\cdot\V{g}^\dagger_{\gamma\delta}
\,,
\end{equation}
where $\V{g}=i\vsigma\sigma_y$ is the vector representation of the spin-triplet. 
Approximating the spectral density for the correlation function with delta peak we have
\begin{equation}
\label{G21}
G^{21}_{\alpha\beta}(Q-p)
= -2\pi i \delta(\omega - \epsilon - \eps_\mathbf{q-p})\big(1-n(\omega - \epsilon) \big)
\delta_{\alpha\beta}.
\end{equation}
Inserting Eq.~\eqref{full-vertex} and Eq.~\eqref{G21} into Eq.~\eqref{sigma12} and integrating out the frequency $\omega$, we obtain
\begin{multline}
\label{Sigma12-final}
\Sigma^{12}_{\alpha \gamma}(p) =  i \int_\V{q} \int_\V{p''} W(\V{q}, \eps_\V{p''}+\eps_\V{q-p''})
n(\eps_{\V{p}''})n(\eps_{\V{q-p''}})
\big(1-n(\eps_\V{p''} + \eps_\V{q-p''} - \eps )  \big) \\
\times \delta(\eps_\V{p''} + \eps_\V{q-p''} - \eps - \eps_{\V{q-p}})
\times \delta_{\alpha\gamma}
\end{multline}
where
%
%
\begin{align}\label{transition-prob}
W(\V{q}, \omega) 
&\ =\  
3\pi\Big(\frac{4\pi}{N(0)}\Big)^2
\sum_m \bigg| 
Y_{1m}(\hat{\V p})
\,{\mathcal C}_m(\V{q},\omega)\,
Y_{1m}^*(\hat{\V p}'')
\bigg|^2 
\,,
\\
{\mathcal C}_m(\V{q},\omega)
&\ 
 =\frac{1}{\vartheta + \xi_m^2 q^2 - i\frac{\pi \omega}{8T}} 
\,.
\end{align}
The internal spin variables were traced out leaving $\Sigma^{12}_{\alpha\gamma} = \Sigma^{12}\,\delta_{\alpha\gamma}$, and yielding a factor of $3=2S+1$ from the three components of the spin-triplet.
The calculation for $\Sigma^{21}$ follows similarly,
\begin{multline}
\Sigma^{21}_{\alpha \gamma}(p) =  -i \int_\V{q} \int_\V{p''} W(\V{q}, \eps_\V{p''}+\eps_\V{q-p''}) 
(1-n(\eps_{\V{p}''}))(1-n(\eps_{\V{q-p''}}))
n(\eps_\V{p''} + \eps_\V{q-p''} - \eps ) \\
\times \delta(\eps_\V{p''} + \eps_\V{q-p''} - \eps - \eps_{\V{q-p}})
\times \delta_{\alpha\gamma}.
\end{multline}
Recall that the right-hand side of the kinetic equation~\eqref{BLE} is
\begin{equation}
\mathrm{RHS} 
=-i\Sigma^{12}\big|_{\eps=\eps_\V{p}}(1-n_{\V{p}}) - i\Sigma^{21}\big|_{\eps=\eps_\V{p}}n_{\V{p}}.
\end{equation}
Inserting the above results into this expression, the right-hand side takes the form of the collision integral,
\begin{multline}
\label{collision}
\mathrm{RHS} =
\int_{\V{q}}\int_{\V{p}'}
W(\eps_\mathbf{p'}+\eps_\mathbf{q-p'}, \mathbf{q})
\Big[ 
n(\eps_\mathbf{p'})n(\eps_\mathbf{q-p'})
\big(1-n(\eps_\mathbf{p'}+\eps_\mathbf{q-p'}-\epsilon_\V{p}) \big)
\big(1-n(\eps_\V{p})\big) \\
-\big(1-n(\eps_\mathbf{p'})\big)\big(1-n(\eps_\mathbf{q-p'})\big)
n(\eps_\mathbf{p'}+\eps_\mathbf{q-p'}-\epsilon_\V{p})
n(\eps_\V{p})
\Big]
\delta(\eps_\mathbf{p'}+\eps_\mathbf{q-p'}-\eps_{\V{p}} - \eps_\mathbf{q-p}).
\end{multline}
In comparison with Emery's phenomenological theory~\cite{eme76}, the above result for the structure of the transition probability in Eq.~\eqref{transition-prob} is a \emph{sum over the probabilities} for quasiparticles scattering from all $(2S+1)(2m+1)=9$ Cooper pair fluctuations, i.e. there are no interference terms. This is to be expected since pair fluctuations with different quantum numbers are \emph{not} phase coherent for $T>T_c$. Emery's heuristic formula for the transition probability assumes that the pair fluctuation amplitudes add coherently, and thus generates a transition probability that includes interference terms.
In addition our result for the collision integral has a smaller prefactor. Emery's collision integral was obtained from an application of Fermi's golden rule with a scattering amplitude representing the coherent superposition of pair fluctuations, while our expression is obtained from the self-energy given by the diagram in Fig.~\refsub{fig-self-energies}{(d)}. 
The golden rule formula can be related to a diagram with two vertices, similar to the second-order diagram in Fig.~\refsub{fig-self-energies}{(c)}, but with the non-singular normal state T-matrix replaced by the vertex $\Gamma$. However, the vertex $\Gamma$ is summed over all repeated scattering in the particle-particle channel, and thus inserting $\Gamma$ into a formula of the form given in diagram Fig.~\refsub{fig-self-energies}{(c)}, as was effectively done by Emery, also double counts the ladder diagrams, which explains why Emery's expression also has a prefactor twice larger than our result, in addition to erroneously adding the pair fluctuation amplitudes coherently.

\begin{figure}[tbp]
\centerline{\includegraphics[width=0.8\textwidth]{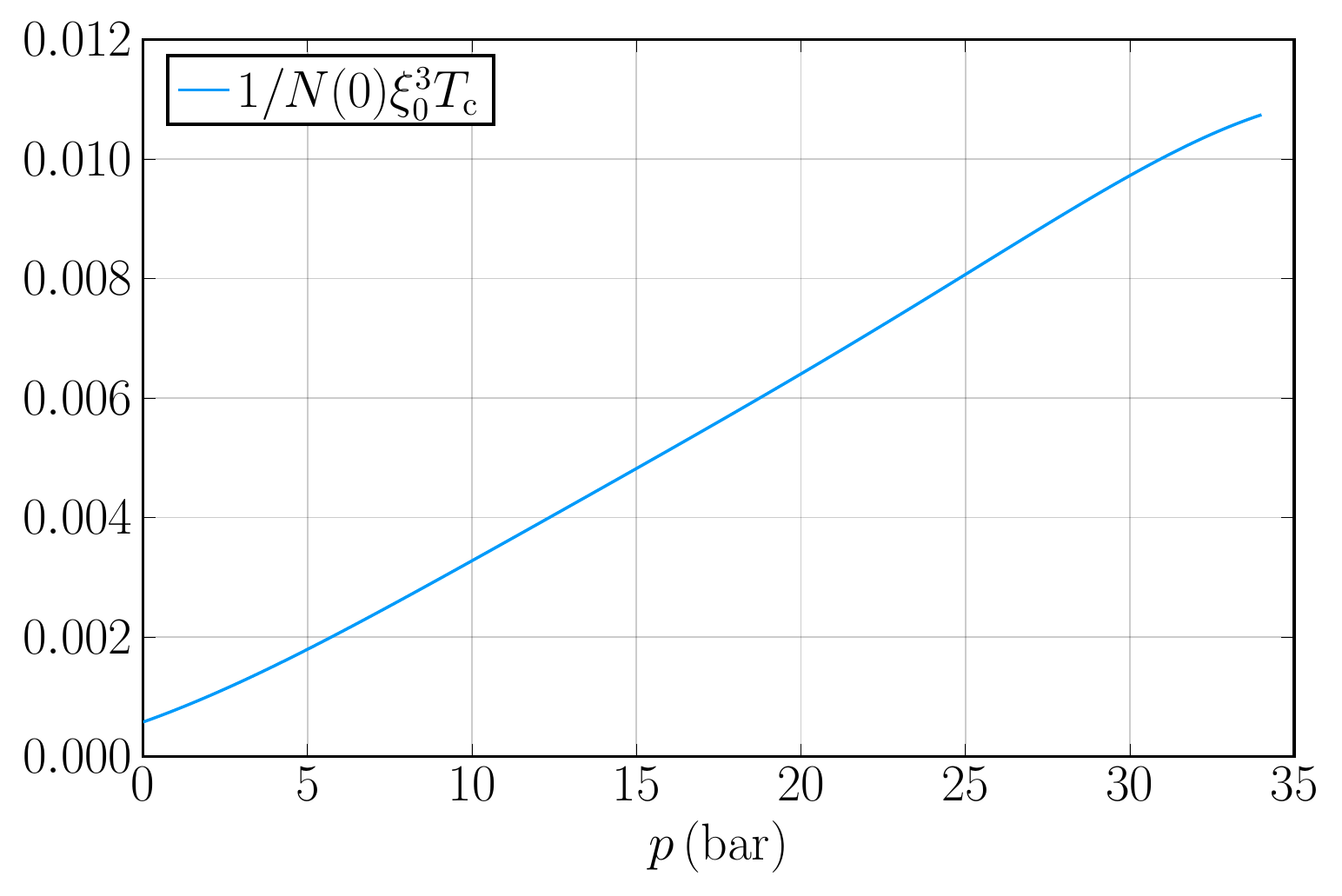}}
\caption[]{Pressure dependence of the dimensionless parameter $\frac{1}{N(0)\xi_0^3 T_{\mathrm{c}}}$ which sets the overall scale for the quasiparticle-pair-fluctuation self-energy.}
\label{fig-N0xiTc}
\end{figure}

The magnitude of the self-energy is formally of order $\sml^3$, and can be estimated as follows. From the expressions \eqref{Sigma12-final}, the integral over Cooper pair momentum $\int d^3q$ gives rise to a dimensional factor $1/\xi_0^3$, and the integral $\int_{\V{p}''} \sim N(0)\int d\eps_{\V{p}''}$ together with a delta function gives a factor $N(0)$.
Recall that the transition probability \eqref{transition-prob} has magnitude $W\sim 1/N(0)^2$.
The overall magnitude for $\Sigma^{12}$ is thus given by
\begin{equation}
\Sigma^{12} \sim \frac{1}{N(0)\xi_0^3} = \frac{T_{\mathrm{c}}}{N(0)\xi_0^3 T_{\mathrm{c}}},
\end{equation}
where the transition temperature $T_{\mathrm{c}}$ gives an estimate of the typical energy scale when the temperature is near $T_{\mathrm{c}}$. The dimensionless parameter $\frac{1}{N(0)\xi_0^3 T_{\mathrm{c}}}$ is a small number, as shown in Figure~\ref{fig-N0xiTc}. The estimate also applies to other self-energies, and thus this parameter controls all phenomena related to leading order corrections to transport from pair fluctuations.
%
%
The magnitude of this prefactor implies the corrections to quasiparticle transport properties due to pair fluctuations are small, except for temperature close to $T_c$. This is consistent with the limited experimental results for propagation and attenuation of zero sound~ \cite{pau78c,lee96}.

\section{Zero Sound Attenuation}\label{sec-zero-sound}

As an application of the nonequilibrium theory of quasiparticle-pair-fluctuation scattering we calculate the zero sound attenuation applicable to liquid \He\ based on the collision integral obtained in Sect.~\ref{sec-collision-integral}. Since the coherence time for zero sound excitations is much longer than the period of zero sound vibrations, we can treat the collision integral as a perturbation to the collisionless kinetic equation that determines the zero sound mode. Adapting the perturbation method of Ref.~\cite{cor69} leads to an attenuation coefficient given by 
\begin{equation}\label{alpha-0}
\alpha_0 = \Big(\frac{-1}{c_0}\Big)
\frac{\displaystyle\int\frac{d\Omega_{\hat{\V{p}}}}{4\pi}
\bigg[
\int d\eps_{\V{p}}\ I_{\V{p}} [\psi^{(0)}]
\bigg]
\psi_{\hat{\V{p}}}^{(0)}
}
{\displaystyle \sum_{l\ge 0}
\frac{(\nu_l^{(0)})^2}{2l+1}
\Big(1+\frac{F_l^s}{2l+1}\Big)
},
\end{equation}
where
\begin{equation}
\label{delta-I}
\begin{split}
I_{\V{p}_1}[\psi^{(0)}] =
& -\beta \int \frac{d^3 p_2}{(2\pi)^3}\int \frac{d^3 p_3}{(2\pi)^3}
W(\V{p}_1, \V{p}_2; \V{p}_3, \V{p}_4) \delta(\eps_1 + \eps_2 - \eps_3 - \eps_4)\\
&\times\Big(\psi_{\hat{\V{p}}_1}^{(0)} + \psi_{\hat{\V{p}}_2}^{(0)}
-\psi_{\hat{\V{p}}_3}^{(0)} - \psi_{\hat{\V{p}}_4}^{(0)}
\Big)
\times \big[\nf(\eps_1)\nf(\eps_2)\big(1-\nf(\eps_3)\big)\big(1-\nf(\eps_4)\big)\big],
\end{split}
\end{equation}
where $\eps_i \equiv \eps_{\V{p}_i}$ and $\V{p}_4 = \V{p}_1 + \V{p}_2 - \V{p}_3$ since we have integrated out the $\V{p}_4$ variable.
The function $\psi_{\hat{\V{p}}}$ is the deviation from the \emph{local} equilibrium, while the function $\nu_{\hat{\V{p}}}$ is the deviation from \emph{global} equilibrium.
They are related by
\begin{equation}
\label{psi-nu-relation}
\psi_{\hat{\V{p}}}=\nu_{\hat{\V{p}}}+\int\frac{d\Omega_{\hat{\V{p}}'}}{4\pi}F^{s}(\hat{\V{p}}\cdot\hat{\V{p}}')\nu_{\hat{\V{p}}'},
\end{equation}
or in terms of spherical harmonics,
\begin{equation}
\label{psi-and-nu}
\psi_l = \nu_l\Big(1 + \frac{F^{s}_l}{2l+1}\Big).
\end{equation}

Equation~\eqref{alpha-0} is a general result so long as the relaxation time is much longer than the period of zero sound.
In the following, we apply this formula to the pair-fluctuation collision integral~\eqref{collision}, where the transition probability is given by the expression~\eqref{transition-prob}.
For the two incoming particles comprising a pair fluctuation, their momenta are roughly opposite, i.e., we can assume the total momentum $\V{q}=0$, except for the $q^2$ dependence in the transition probability $W$ given in Eq.~\eqref{transition-prob}. 
Hence the two particles have the same energy, and the collision integral~\eqref{delta-I} becomes 
\begin{multline}
I^{\mbox{\tiny C}}_{\V{p}} =
-\beta \int \frac{d^3 q}{(2\pi)^3}\int \frac{d^3 p'}{(2\pi)^3}
W(\V{q}, 2\eps_{\V{p}})
\Big(\psi_{\hat{\V{p}}}^{(0)} + \psi_{-\hat{\V{p}}}^{(0)}
-\psi_{\hat{\V{p}}'}^{(0)} - \psi_{-\hat{\V{p}}'}^{(0)}
\Big) \\
\times\delta(2\eps_{\V{p}} - 2\eps_{\V{p}'}) \times
\nf(\eps_{\V{p}})^2\big(1-\nf(\eps_{\V{p}'})\big)^2,
\end{multline}
where we have changed the integral variable from $\V{p}_2$ to $\V{q}=\V{p}_2+\V{p}_1$.\footnote{The momenta $\V{p}_1$, $\V{p}_2$, $\V{p}_3$, and $\V{p}_4$ in the above formula correspond to $\V{p}$, $\V{q-p}$, $\V{p}'$, and $\V{q-p'}$ here, respectively.}
By carrying out the energy integral $\int d\eps_{\V{p}'}$, we obtain
\begin{multline}
I^{\mbox{\tiny C}}_{\V{p}} =
-\beta \int \frac{d^3 q}{(2\pi)^3}
N(0)\int \frac{d\Omega_{\hat{\V{p}}'}}{4\pi}
W(\V{q}, 2\eps_1)
\Big(\psi_{\hat{\V{p}}}^{(0)} + \psi_{-\hat{\V{p}}}^{(0)}
-\psi_{\hat{\V{p}}'}^{(0)} - \psi_{-\hat{\V{p}}'}^{(0)}
\Big)  \\
\times\frac{1}{2}
\nf(\eps_{\V{p}})^2\big(1-\nf(\eps_{\V{p}})\big)^2.
\end{multline}
The $\psi_{\hat{\V{p}}}^{(0)}$ are collisionless zero sound eigenfunctions, and thus can be expanded in Legendre polynomials, $P_l(\hat{\V{p}}\cdot\hat{\V{k}})$, where $\V{k}=k\hat{\V{k}}$ is the zero sound wavevector, 
\begin{equation}
\psi_{\hat{\V{p}}}^{(0)} = \sum_{l\ge 0} \psi_l^{(0)} P_l(\hat{\V{p}}\cdot\hat{\V{k}}).
\end{equation}
We can neglect the components with $l\ge 3$~\cite{bay78}, and the $l=0,1$ components drop out because of particle number conservation and momentum conservation, thus
\begin{equation}
\Big(\psi_{\hat{\V{p}}}^{(0)} + \psi_{-\hat{\V{p}}}^{(0)}
-\psi_{\hat{\V{p}}'}^{(0)} - \psi_{-\hat{\V{p}}'}^{(0)}
\Big) 
= 2\psi_2^{(0)} \big[P_2(\hat{\V{p}}\cdot\hat{\V{k}}) - P_2(\hat{\V{p}}'\cdot\hat{\V{k}})\big] .
\end{equation}
From the numerator in the first-order solution~\eqref{alpha-0}, we have
\begin{equation}
\mathrm{Num} =\int\frac{d\Omega_{\hat{\V{p}}}}{4\pi}
\bigg[
\int d\eps_{\V{p}}\ I^{\mbox{\tiny C}}_{\V{p}} [\psi^{(0)}]
\bigg]
\psi_{\hat{\V{p}}}^{(0)} .
\end{equation}
Inserting the above expression for $I^{\mbox{\tiny C}}_{\V{p}}$ into this formula, we have
\begin{multline}
\mathrm{Num} =
-\beta N(0) \times 2\psi_2^{(0)}
\int d\eps_{\V{p}}  \int\frac{d\Omega_{\hat{\V{p}}}}{4\pi} \ 
\psi_{\hat{\V{p}}}^{(0)}
\int \frac{d^3 q}{(2\pi)^3}
\int \frac{d\Omega_{\hat{\V{p}}'}}{4\pi} \ 
W(2\eps_{\V{p}},\V{q})
\big[P_2(\hat{\V{p}}\cdot\hat{\V{k}}) - P_2(\hat{\V{p}}'\cdot\hat{\V{k}})\big] \\
\times \frac{1}{2}
\nf(\eps_{\V{p}})^2\big(1-\nf(\eps_{\V{p}})\big)^2 .
\end{multline}
The angular variables $\hat{\V{q}}$, $\hat{\V{p}}'$, and $\hat{\V{p}}$ can then be integrated out easily.\footnote{Note that the spherical harmonics in the expression \eqref{transition-prob} is defined with respect to the direction $\hat{\V{q}}$, while the expansion of $\psi^{(0)}_{\hat{\V{p}}}$ is relative to the direction of zero sound wavevector $\hat{\V{k}}$. We do the $\hat{\V{q}}$ integral first, and then $\hat{\V{p}}'$ and $\hat{\V{p}}$. Since the zero sound mode is degenerate with respect to $\hat{\V{k}}$, the result is independent of $\hat{\V{k}}$.}
We obtain
\begin{equation}\label{num-integral}
\mathrm{Num} =
-\beta \frac{27\pi}{N(0)} \frac{(\psi_2^{(0)})^2}{5}
\int d\eps_{\V{p}}
\int \frac{q^2 dq}{2\pi^2}
\Big[\frac{7}{75}|\mathcal{C}_0|^2 
+ 2\times\frac{1}{4}\frac{32}{75}|\mathcal{C}_1|^2 \Big]
\nf(\eps_{\V{p}})^2\big(1-\nf(\eps_{\V{p}})\big)^2,
\end{equation}
where 
\begin{equation}\label{eq-Cooper_pair_propagator}
\mathcal{C}_m \equiv \frac{1}{\vartheta + \xi_{1m}^2 q^2 - i\frac{\pi 2\eps_{\V{p}}}{8T}},\quad m=0,\pm 1
\end{equation}
is the singular Cooper pair fluctuation amplitude for pairs in orbital state $Y_{1m}(\hat{\V{p}})$.
The denominator of Eq.~\eqref{alpha-0} is 
\begin{equation}
\mathrm{Denom} = 
\sum_{l\ge 0}
\frac{(\nu_l^{(0)})^2}{2l+1}
\Big(1+\frac{F_l^{s}}{2l+1}\Big)
\,,
\end{equation}
which good approximation can be shown to be (c.f. App.~\ref{sec-denom}), 
\begin{equation}\label{denom}
\mathrm{Denom} = 3\nu_0^2\frac{m}{m^*} 2s_0^2
\,.
\end{equation}

\begin{figure}[tbp]
\centerline{\includegraphics[width=0.9\textwidth]{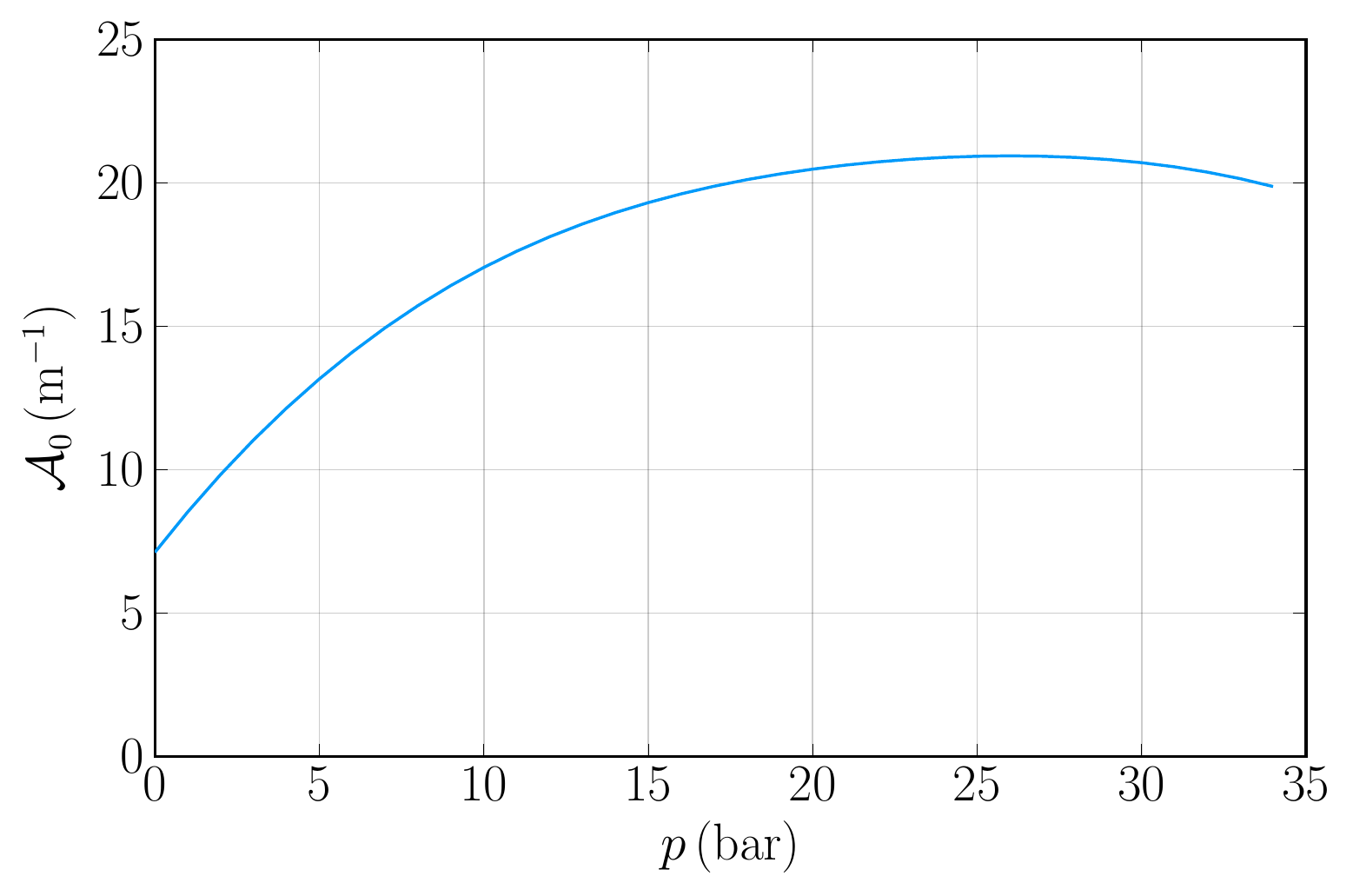}}
\caption{Pressure dependence of the prefactor $\mathcal{A}_0$ for the excess attenuation from quasiparticle-pair-fluctuation scattering.}
\label{fig-attenuation-pressure}
\end{figure}

The double integral in Eq.~\eqref{num-integral}, which has a singular integrand for $q\rightarrow 0$, $\eps_1\rightarrow 0$ and $\vartheta\rightarrow 0^+$, is integrable.
In particular the occupation factor $n_{\fm}(\eps_1)(1-n_{\fm}(\eps_1))$ is exponentially small for $|\beta\eps_1|\gg 1$, which ensures convergence. Carrying out the integration over $q$ first in Eq.~\eqref{num-integral} yields
\begin{equation}
\int_0^{\infty} q^2 dq |\mathcal{C}_m|^2 = 
\frac{1}{\xi_{1m}^3} \frac{\pi}{2} \frac{\mathfrak{Im}\sqrt{\vartheta + i|x|}}{|x|}, 
\end{equation}
where we scaled the remaining energy integration with $x\equiv 2\pi\eps_1/8T$. The occupation factor can then be written as $n_{\fm}(\eps_1)(1-n_{\fm}(\eps_1))=\cosh^{-2}(\eps_1/2T)/4=\sech^{2}(2x/\pi)/4$. 
All three orbital contributions are proportional to the same function of $x$, and thus a single dimensionless function of the reduced temperature,  
\begin{equation}\label{Theta}
\Theta(\vartheta) 
\equiv
\frac{1}{2}\int_{-\infty}^{\infty}dx\,\frac{\mathfrak{Im}\sqrt{\vartheta + i|x|}}{|x|}
                                      \times \sech^4(2x/\pi)
\,,
\end{equation}
which is calculated numerically and is finite at $T_c$, i.e. $\vartheta=0$. However $\Theta(0)$ includes contributions to the attenuation from short wavelengths, $q\gg \pi/\xi_0$, and thus beyond the region of validity of the long-wavelength expression for $\mathcal{C}_{m}(\V{q},\omega)$. Thus, we introduce a counter term, $\alpha^{\text{C}}_{\infty}$, to correct for the short wavelength contributions to $\Theta(\vartheta)$ based on Eq.~\eqref{eq-Cooper_pair_propagator}. 
The result for the quasiparticle-pair-fluctuation contribution to the attenuation of zero sound becomes
\begin{equation}\label{attenuation-no-cutoff-2}
\alpha_0^{\mbox{\tiny C}} 
= \frac{1}{10\pi} \frac{1}{c_0 s_0^2}
\frac{m^{*}}{m}
\frac{1}{N(0)}
\Big(1 + \frac{F^{s}_2}{5}\Big)^2
\frac{N_{\#}}{\xi_0^3}
\Theta(\vartheta) + \alpha_{\infty}^{\mbox{\tiny C}}
\,.
\end{equation}
The counter term is then determined by the measured value of the excess attenuation at $T_c$, and the number $N_{\#}$ is given by
\begin{equation}\label{eq-Number}
N_{\#}=\frac{9}{2}\bigg[\frac{7}{75}\sqrt{\frac{5}{9}}^3 + 2\frac{8}{75}\sqrt{\frac{5}{3}}^3\bigg]
\approx 2.24
\,.
\end{equation}
This result follows from the sum over modes with $\xi_\parallel^2 = \frac{9}{5}\xi_0^2$ and $\xi_{\pm 1}^2\equiv\xi_\perp^2 = \frac{3}{5}\xi_0^2$, where $\xi_0^2=(7\zeta(3)/48\pi^2)(v_f/T_c)^2$.
%
\begin{figure}[tbp]
\centerline{\includegraphics[width=1.0\textwidth]{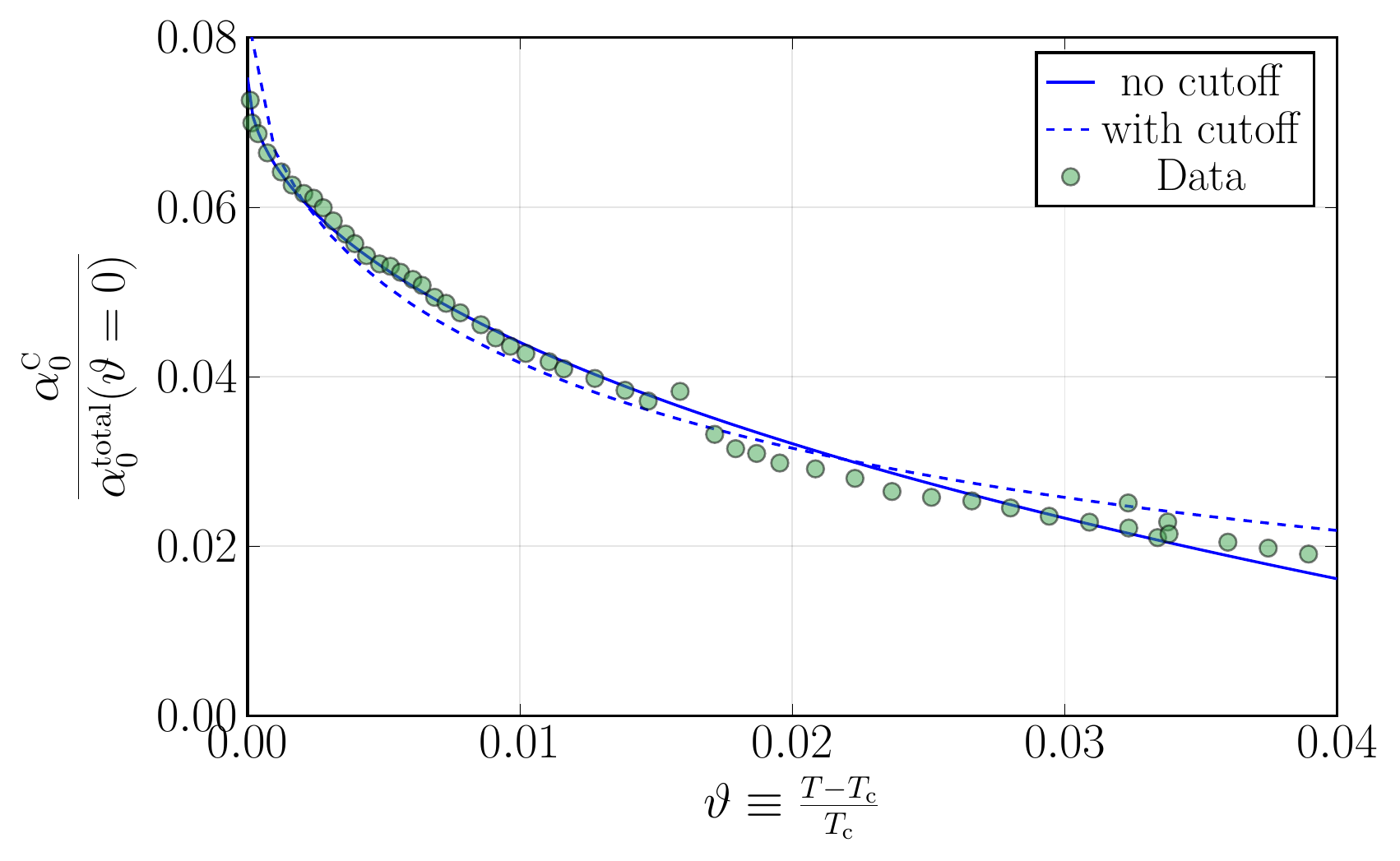}}
\caption{Excess attenuation normalized to the total attenuation at $T_c$. The data are from Ref.~\cite{pau78c}. The solid blue curve is our theoretical prediction from Eq.~\eqref{attenuation-no-cutoff-2}, with the counter term and $F^s_2$ determined by the best fit of the excess attenuation over full experimental range of the data, which yields $\alpha^{\mbox{\tiny C}}_{\infty} = -0.215\times\alpha_0^{\text{total}}(\vartheta=0)$ and $F^s_2=1.39$. The dashed blue curve is the best fit using the approximate formula in Eq.~\eqref{attenuation-cutoff} with $x_{\mathrm{c}}=0.236$ and $F^{s}_2 = 3.25$.}
\label{fig-attenuation-fit}
\end{figure}
%
The magnitude of the temperature-dependence of the excesss attenuation is determined by the prefactor of $\Theta(\vartheta)$ in Eq.~\eqref{attenuation-no-cutoff-2}, which can be expressed as
\begin{equation}\label{prefactor}
\begin{split}
\mathcal{A}_0
&= \frac{2.24}{10\pi}\frac{1}{N(0)\xi_0^3 T_{\mathrm{c}}}
\frac{m^*}{m}\Big(\frac{v_\fm}{c_0}\Big)^3 
\left(1+\frac{F_2^{s}}{5}\right)^2
\frac{T_{\mathrm{c}}}{v_\fm}.
\end{split}
\end{equation}
Figure \ref{fig-attenuation-pressure} shows the increase in the prefactor $\mathcal{A}_0$ with pressure based on the Fermi liquid data from Ref.~\cite{har00} and the value of $F_2^s=1.39$ obtained from optimizing the fit of Eq.~\eqref{attenuation-no-cutoff-2} to the data of Paulson and Wheately~\cite{pau78c}.
The predicted excess attenuation from quasiparticle-pair-fluctuation scattering is a few percent of the total attenuation at $T_c$, which ranges from $150\,\mathrm{m}^{-1}$ to $250\,\mathrm{m}^{-1}$~\cite{wat03} over the pressure range from $0$ to $34\,\mathrm{bars}$.
The excess attenuation reported by Paulson and Wheatley~\cite{pau78c} ranges from 2--8\% percent (largest at melting pressure) of the baseline attenuation from Fermi liquid theory, consistent with the $\cO(\sml^3)$ contribution from the quasiparticle-pair-fluctuation self-energy term. In particular, the pressure dependence of the prefactor $\mathcal{A}_0$ shown in Fig.~\ref{fig-attenuation-pressure} is consistent with the report of Paulson and Wheatley~\cite{pau78c}.
Last, but not least, the temperature dependence of the excess attenuation is in excellent agreement with the prediction of Eqs.~\eqref{Theta},~\eqref{attenuation-no-cutoff-2} and \eqref{eq-Number}, as shown in Fig.~\ref{fig-attenuation-fit} (solid blue line), providing validation of the theory of quasiparticle-pair-fluctuation scattering.

\section{Cutoff Procedure of Samalam and Serene}\label{sec-zero-sound-cutoff}

Motivated by the experimental report of an excess attenuation of zero sound near $T_c$~\cite{pau78c}, Samalam and Serene~\cite{sam78} calculated the attenuation of zero sound in \He\ based on Emergy's heuristic collision integral. Their result for the temperature dependence of the attenuation is based on a cutoff procedure to exclude pair-fluctuations with wavevectors $q \gtrsim \xi_0^{-1}$. This cutoff procedure leads to an analytic formula for the excess attenuation that depends on $\sqrt{\vartheta}$ and the cutoff. 

Here we apply the cutoff procedure of Samalam and Serene to evaluate the double integral in Eq.~\eqref{num-integral} for our result for the transition probability for scattering quasiparticles from pair fluctuations, and compare this result with our renormalization procedure. 

Starting from the double integral in Eq.~\ref{num-integral}
\begin{equation}\label{double_integral}
J = \int_{-\infty}^{+\infty} d\eps_{\V{p}}\,\left[n_F(\eps_{\V{p}})(1-n_F(\eps_{\V{p}})\right]^2
    \int_0^{\infty}q^2 dq\,\left|\mathcal{C}_m(q,2\eps_{\V{p}})\right|^2
\,,
\end{equation}
and following Samalam and Serene, we approximate the Fermi distribution functions in Eq.~\eqref{double_integral} by their values at $\eps_{\V{p}}=0$, i.e. $n_F(0)=(1-n_F(0)) = \nicefrac{1}{2}$, then carry out the integration over $\eps_{\V{p}}$ by closing the contour in the upper half plane to obtain,
\begin{equation}\label{approx}
\int_{-\infty}^{+\infty}d\eps_{\V{p}} |\mathcal{C}_m(q,2\eps_{\V{p}})|^2 = \frac{4T}{\vartheta + \xi_m^2 q^2}
\,.
\end{equation}

The remaining phase space integration over the pair-fluctuation modes is now ultraviolet divergent. Samalam and Serence introduce an ultraviolet cutoff to eliminate modes with $\xi_m q \ge x_c$, with the presumption that $x_c\sim\mathcal{O}(1)$. The resulting integral over modes gives,
\begin{equation}\label{q-integral}
\int_0^{q_c} dq\frac{q^2}{\vartheta + \xi_{1m}^2 q^2}
= 
\frac{1}{\xi_{1m}^3} \int_0^{x_\mathrm{c}} dx \frac{x^2}{\vartheta + x^2} \\
= 
\frac{1}{\xi_{1m}^3} \Big[x_\mathrm{c} - \sqrt{\vartheta}\arctan(x_\mathrm{c}/\sqrt{\vartheta}) \Big]
\,.
\end{equation}
The rest of the calculation for the excess attenuation is the same as described in the previous section yielding the the following analytic approximation for the pair-fluctuation contribution to the zero sound attenuation,

\begin{equation}\label{attenuation-cutoff}
\alpha_0^{\mbox{\tiny C}} 
\simeq 
\mathcal{A}_0\,\times\,
\Big[x_\mathrm{c}-\sqrt{\vartheta}\arctan(x_\mathrm{c}/\sqrt{\vartheta})\Big]
\,.
\end{equation}
N.B.\ there is no counter term since the short-wavelength modes have been excluded by the cutoff. 


In Fig.~\ref{fig-attenuation-fit} our result based on the cutoff procedure of Samalam and Serene is shown in comparison with our renormalization procedure. While the latter (Eq.~\eqref{attenuation-no-cutoff-2}) gives a more accurate result for the temperature dependence of the excess attenuation, the cutoff procedure (Eq.~\eqref{attenuation-cutoff}) also provides a reasonable representation of the experimental data obtained by Paulson and Wheatly~\cite{pau78c} with the least square fit yielding $x_{\mathrm{c}}\approx 0.236$ and $F^{s}_2\approx 3.25$. 
For comparison, our result based on Eq.~\eqref{attenuation-no-cutoff-2}, gives from the best fit $\alpha_{\infty}^{\mbox{\tiny C}}\approx -0.215\times\alpha_0^{\text{total}}(\vartheta=0)$ and $F^{s}_2\approx 1.39$.
We do not have precise data for the value of $F^{s}_2$, but the smaller value of $F^{s}_2$ given by our renormalization procedure is close to currently reported experimental results.\footnote{See the summary plot given by Halperin's group in \url{http://spindry.phys.northwestern.edu/3He Calculator/F2s_plot.html}, and Sect.~5.4 of the book by Dobbs~\cite{dobbs00}.}

Nevertheless, the cutoff procedure of Samalam and Serene has the advantage of analytic results for the temperature dependence of the pair-fluctuation corrections, and is sufficiently accurate in the case of the excess zero-sound attenuation that we use this procedure below to calculate the pair-fluctuation corrections to the Fermi velocity and Fermi liquid properties that are derived from the real part of the quasiparticle-pair-fluctuation self-energy (diagram~\refsub{fig-self-energies}{(d)}). In the following section we calculate the corrections to Fermi velocity and Fermi liquid parameters, and in Ref.~\cite{lin21a} we report the pair-fluctuation corrections to the zero sound velocity for liquid \He.

\section{Fermi Liquid Parameters}\label{sec-param}

The corrections to the Fermi velocity and Fermi liquid parameters result from the contributions to the left-hand side of Eq.\eqref{BLE} from the $\Sigma^{11(d)}$. Thus, we consider just the collisionless Boltzmann-Landau equation with pair-fluctuation corrections,
\begin{multline}
\label{LHS-of-BLE}
(\partial_t + \frac{\V{p}}{m}\cdot \nabla_\V{R})n_{\V{p}}
- \nabla_\V{R}\mathfrak{Re}\Sigma^{11}\big|_{\eps=\eps_\V{p}}\cdot\nabla_\V{p}n_{\V{p}}
+ \nabla_\V{p}\mathfrak{Re}\Sigma^{11}\big|_{\eps=\eps_\V{p}}\cdot\nabla_\V{R}n_{\V{p}} 
- \partial_\eps \mathfrak{Re}\Sigma^{11}\big|_{\eps=\eps_\V{p}} \partial_t n_{\V{p}} 
=0.
\end{multline}
The real part of the self-energy $\Sigma^{11}$ determines the Fermi velocity (or equivalently, the effective mass), and the Landau parameters. As can be shown by the quasiclassical method~\cite{ser83}, the dominating contribution to these parameters comes from the molecular field. When the temperature is near $T_{\mathrm{c}}$, the contribution from the diagram in Figure~\refsub{fig-self-energies}{(d)}, which contains the effect of Cooper pair fluctuations, becomes strong. In this section, we calculate the corrections from the fluctuation self-energy by considering the equation \eqref{LHS-of-BLE}. We first separate the equation into two parts. The first part comes from the molecular field, giving the standard Boltzmann-Landau equation. The second part is given by the fluctuation self-energy. Thus the equation can be written as 
\begin{multline}
\big(1-\partial_\eps \mathfrak{Re}\Sigma_{\mathrm{fluc}}^{11}\big|_{\eps=\eps_\V{p}}\big)\partial_tn_{\V{p}}
+ 
\big(\V{v}_{\V{p}}+\nabla_\V{p}\mathfrak{Re}\Sigma_{\mathrm{fluc}}^{11}
\big|_{\eps=\eps_\V{p}}\big)\cdot\nabla_\V{R}n_{\V{p}} 
-
\nabla_\V{R}\big(\delta\eps_{\V{p}}+\mathfrak{Re}\Sigma_{\mathrm{fluc}}^{11}
\big|_{\eps=\eps_\V{p}}\big)\cdot\nabla_\V{p}n_{\V{p}}
=0,
\end{multline}
where the velocity $v_{\mathrm{F}}\equiv\V{p}/m^{*}=\partial_{\V{p}}\eps^0_{\V{p}}$ and the energy from the nonequilibrium distribution is given by
\begin{equation}
\delta\eps_{\V{p}} = 
\frac{1}{V} \sum_{\V{p}'\sigma'} f_{\V{p}\sigma, \V{p}'\sigma} \delta n_{\V{p}'\sigma'}.
\end{equation}
Note that the coefficient of the time derivative term is also modified. Assume the deviation from equilibrium is weak and the kinetic equation can be linearized. To the leading order we have
\begin{multline}
\label{linear-BLE} 
\big(1 - \partial_\eps \mathfrak{Re}\Sigma_{\mathrm{fluc}}^{11}\big|_{\eps=\eps_\V{p}}^{\mathrm{eq}}\big)  \partial_t\delta n_{\V{p}}
+ \big(\V{v}_{\V{p}} + \nabla_\V{p}\mathfrak{Re}\Sigma_{\mathrm{fluc}}^{11}\big|_{\eps=\eps_\V{p}}^{\mathrm{eq}}\big)\cdot \nabla_\V{R}\delta n_{\V{p}} 
- \nabla_\V{R}\big(\delta\eps_{\V{p}}+\mathfrak{Re}\delta\Sigma_{\mathrm{fluc}}^{11}\big|_{\eps=\eps_\V{p}}\big)\cdot\nabla_\V{p}n_{\V{p}}^{\mathrm{eq}}
=0,
\end{multline}
In the following we drop the subscript ``fluc''. Recall that $\mathfrak{Re}\Sigma^{11} = (\Sigma^{11} - \Sigma^{22})/2$. Using 
\begin{align}
\Sigma^{11}(p)&=\sum_Q \Gamma^{11}_{\alpha\beta,\gamma\delta}(p,p;Q)G^{11}_{\beta\delta}(Q-p) \\
\Sigma^{22}(p)&=\sum_Q \Gamma^{22}_{\alpha\beta,\gamma\delta}(p,p;Q)G^{22}_{\beta\delta}(Q-p)
\end{align}
and
\begin{align}
\Gamma^{11} &= \sum_\lambda \hat{p}_i \frac{-L_\lambda^{\mathrm{GL}} - (GG)^{(21)}_\lambda}
{|L^\mathrm{GL}_\lambda|^2}
P^\lambda_{ij} \hat{p}_j' \\
\Gamma^{22} &= \sum_\lambda \hat{p}_i \frac{L_\lambda^{\mathrm{GL}} - (GG)^{(12)}_\lambda}
{|L^\mathrm{GL}_\lambda|^2}
P^\lambda_{ij} \hat{p}_j'
\end{align}
and $G^{11*}(p) = -G^{22}(p)$, we obtain
\begin{multline}
\label{real-part}
\frac{\Sigma^{11}-\Sigma^{22}}{2} = 
\frac12 \int_{Q} \sum_\lambda \Big[ \hat{p}_i \frac{-(GG)^{21}_\lambda - (GG)^{12}_\lambda}{|L^\mathrm{GL}_\lambda|^2}
P^\lambda_{ij} \hat{p}_j \Big]
\mathfrak{Re}G^{11}(Q-p) \\ 
+\frac12 \int_{Q} \sum_\lambda \Big[ \hat{p}_i \frac{(L^\mathrm{GL}_\lambda)^* - L^\mathrm{GL}_\lambda}{|L^\mathrm{GL}_\lambda|^2}
P^\lambda_{ij} \hat{p}_j \Big]
i\mathfrak{Im}G^{11}(Q-p).
\end{multline}
We evaluate this quantity in equilibrium for the time derivative and the Fermi velocity, and in nonequilibrium for the Landau parameters.

\subsection{Fermi velocity and the time derivative}
The self-energy term for the Fermi velocity and the time derivative in \eqref{linear-BLE} is calculated in equilibrium.
The relevant expressions can be found in Section \ref{sec-fluctuation}.
Recall that in equilibrium
\begin{equation}
(GG)^{12}=
-\pi \frac{N(0)}{3}
\Big[
\Big(\frac{1}{4} - \frac{\omega}{8T}\Big)\delta_{ik} 
-\frac{3\pi^2}{4\times 5 \times 7\zeta(3)}
\xi_0^2 q^2
(\delta_{ik}+2\hat{q}_i \hat{q}_k)
\Big]
\end{equation}
and using $(GG)^{12}(q) = (GG)^{21}(-q)$, we have
\begin{equation}
\begin{split}
(GG)^{12}_{ij} + (GG)^{21}_{ij}
&= -\pi \frac{N(0)}{3}
\Big[ \frac{1}{2}\delta_{ij} - \frac{1}{2\times 16\times 5} \frac{v_\mathrm{F}^2 q^2}{T^2}
(\delta_{ij} + 2\hat{q}_i \hat{q}_j)
\Big] \\
&= -\pi \frac{N(0)}{3}
\Big[ \frac{1}{2}(P^\parallel_{ij} + P^\perp_{ij}) - \frac{3\pi^2}{2\times 5 \times 7\zeta(3)}
\xi_0^2 q^2 (3P^\parallel_{ij} + P^\perp_{ij})
\Big].
\end{split}
\end{equation}
The above expression can be separated into the two components $\lambda =\parallel, \perp$.
We have
\begin{equation}
(GG)^{12}_{\lambda} + (GG)^{21}_{\lambda}
= -\pi \frac{N(0)}{6}
\Big[ 1 -
\frac{\pi^2}{7\zeta(3)}\xi_\lambda^2 q^2  \Big]
\end{equation}
with $\xi_\parallel^2 = \frac{9}{5}\xi_0^2$ and $\xi_\perp^2 = \frac{3}{5}\xi_0^2$,
\begin{equation}
-L^\mathrm{GL}_\lambda + (L^\mathrm{GL}_\lambda)^* = -2i\frac{N(0)}{3}(\vartheta + \xi_\lambda^2 q^2),
\end{equation}
and the denominator is
\begin{equation}
|L_{\lambda}^{\mathrm{GL}}|^{2}=\frac{N(0)^{2}}{9}\Big[(\vartheta+\xi_{\lambda}^{2}q^{2})^{2}+\big(\frac{\pi\omega}{8T}\big)^{2}\Big],
\end{equation}
which contains the singular part.
Using the relation $G^{11} = G^{\mathrm{R}} + G^{12}$,
the time-ordered Green function can be written as
\begin{equation}
\label{G11}
G^{11}(q-p)
  =\mathrm{P}\frac{1}{\omega-\epsilon-\epsilon_{\mathbf{q-p}}}  \\
  +i\pi\delta(\omega-\epsilon-\epsilon_{\mathbf{q-p}})[2n(\omega-\epsilon)-1].
\end{equation}
From the equation \eqref{linear-BLE}, the correction is $\nabla_\V{p}\mathfrak{Re}\Sigma_{\mathrm{fluc}}^{11}\big|_{\eps=\eps_\V{p}}^{\mathrm{eq}}$.
We can write the derivative as $\V{v}_{\V{p}}\partial_{\eps_{\V{p}}}$.
Plugging these expressions into \eqref{real-part} for $\mathfrak{Re}\Sigma_{\mathrm{fluc}}^{11}$ and taking the derivative give the result
\begin{multline}
\label{vf-correction}
  \frac{\delta v_{\mathrm{F}}}{v_{\mathrm{F}}}=
  -\frac{4.7174}{2^{10}}\frac{1}{N(0)T_{\mathrm{c}}\xi_{0}^{3}}\times
  \Big[\frac{\arctan\frac{x_{\mathrm{c}}}{\sqrt{\vartheta}}}{\sqrt{\vartheta}^{3}}+
  \frac{x_{\mathrm{c}}^{3}-\vartheta x_{\mathrm{c}}}{\vartheta x_{\mathrm{c}}^{4}+2\vartheta^{2}x_{\mathrm{c}}^{2}+\vartheta^{3}} \\
  -\frac{\pi^{2}}{7\zeta(3)}\Big(\frac{3\arctan\frac{x_{\mathrm{c}}}{\sqrt{\vartheta}}}{\sqrt{\vartheta}}-
  \frac{5x_{\mathrm{c}}^{3}+3\vartheta x_{\mathrm{c}}}{x_{\mathrm{c}}^{4}+2\vartheta x_{\mathrm{c}}^{2}+\vartheta^{2}}\Big)\Big]\\
  -5.975\times10^{-2}\times\frac{1}{N(0)T\xi_{0}^{3}}\Big[x_{\mathrm{c}}-\sqrt{\vartheta}\arctan\frac{x_{\mathrm{c}}}{\sqrt{\vartheta}}\Big],
\end{multline}
where we have taken $\eps_{\V{p}}=0$.
This correction is always negative, so the fluctuations reduce the Fermi velocity. 

\begin{figure}[tbp]
\centering
\begin{subfigure}[b]{0.49\textwidth}
\centering
\includegraphics[width=1.0\textwidth]{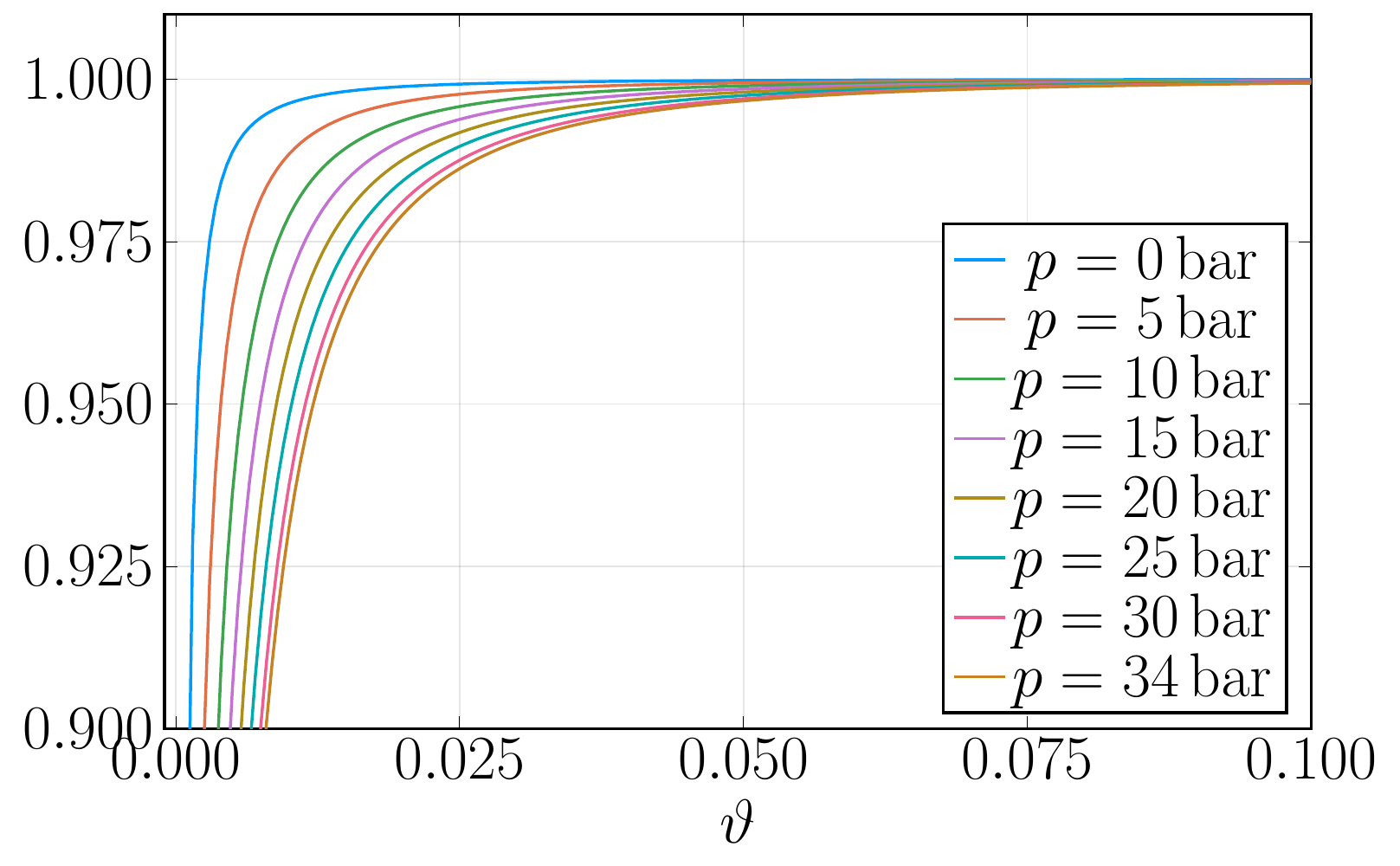}
\caption{}
\end{subfigure}
\hfill
\begin{subfigure}[b]{0.49\textwidth}
\centering
\includegraphics[width=1.0\textwidth]{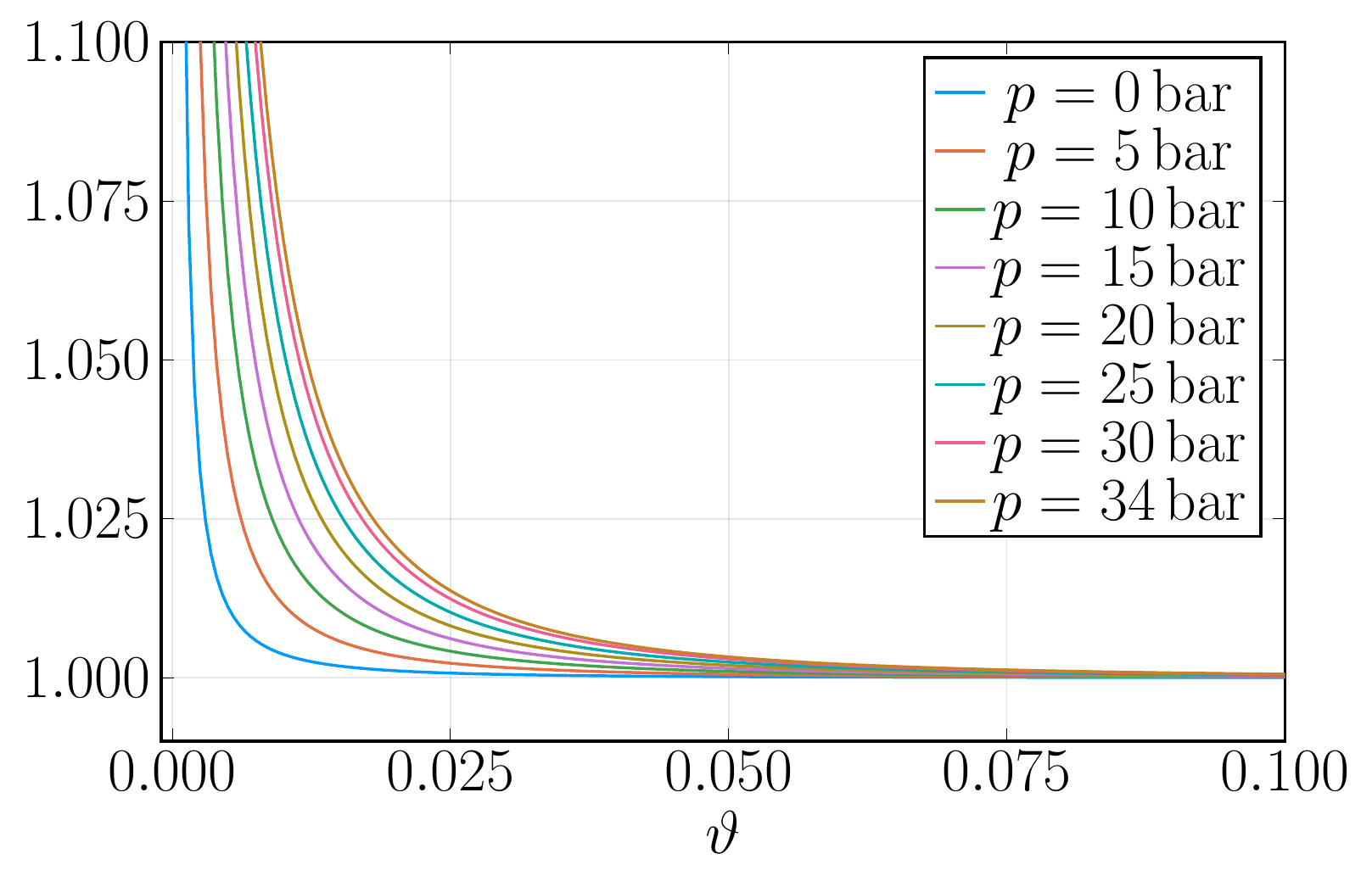}
\caption{}
\end{subfigure}
\caption{Corrections to the Fermi velocity and the constant in front of the time derivative. Panel~(a) gives $1+\delta v_\mathrm{F}/v_\mathrm{F}$, using the formula \eqref{vf-correction} with different pressures, and Panel~(b) gives $1+\delta$, with the correction $\delta$ given by \eqref{dt-correction}. The cutoff is $x_\mathrm{c}=0.236$.}
\label{fig-fermi-velocity}
\end{figure}

The correction to the constant in front of the time derivative (see \eqref{linear-BLE}) can be calculated in a similar way.
The result is
\begin{multline}
\label{dt-correction}
\delta = 
\frac{4.7174}{2^{10}}\frac{1}{N(0)T_{\mathrm{c}}\xi_{0}^{3}}\times\Big[\frac{\arctan\frac{x_{\mathrm{c}}}{\sqrt{\vartheta}}}{\sqrt{\vartheta}^{3}}+\frac{x_{\mathrm{c}}^{3}-\vartheta x_{\mathrm{c}}}{\vartheta x_{\mathrm{c}}^{4}+2\vartheta^{2}x_{\mathrm{c}}^{2}+\vartheta^{3}} 
-\frac{\pi^{2}}{7\zeta(3)}\Big(\frac{3\arctan\frac{x_{\mathrm{c}}}{\sqrt{\vartheta}}}{\sqrt{\vartheta}}-\frac{5x_{\mathrm{c}}^{3}+3\vartheta x_{\mathrm{c}}}{x_{\mathrm{c}}^{4}+2\vartheta x_{\mathrm{c}}^{2}+\vartheta^{2}}\Big)\Big]
\end{multline}
and the time derivative becomes $(1+\delta)\partial_{t}$.
The results for the Fermi velocity and the time constant are shown in Figure~\ref{fig-fermi-velocity} for different pressures, using the cutoff $x_{\mathrm{c}}=0.236$, which is a reasonable estimate obtained from the Paulson and Wheatley's data~\cite{pau78c} in Section~\ref{sec-collision-integral}.
%
%

\subsection{Landau parameters}
The corrections to Landau parameters depend on the nonequilibrium distribution. 
From the expressions in \eqref{GG11}--\eqref{GG22} and \eqref{LGL}, we see that the real part of $\Sigma^{11}$ in Eq.~\eqref{real-part} depends on the distribution function in a complicated way.
To simplify the calculation, we keep the denominator the same as its equilibrium value, $L^{\mathrm{GL}}_{\lambda}\sim (\vartheta + \xi_\lambda^2 q^2 - i\pi\omega/8T)$, which gives the singular contribution as $\vartheta\to 0^{+}$.
This is the same as what we do in the calculation of the collision integral, and is justified for linear response.
Moreover, we assume that the nonequilibrium affects the system mainly through the explicit dependence of the distribution function (i.e. through the $G^{12}$ and $G^{21}$ lines in the diagrams), and the implicit dependence involved in the quasiparticle energy $\eps_\V{p}$ has little influence on the ladder diagrams.

Consider the first term in the right-hand side of Eq.~\eqref{real-part}.
The real part of the time-order Green function \eqref{G11} has no explicit dependence on the distribution function, and thus we only need to consider
\begin{equation}
\hspace{-1em}
  (GG)_{ij}^{21}+(GG)_{ij}^{12}=
  -2\pi\int_{\V{p}'}\delta(\omega-\epsilon_{\V{p}'}-\epsilon_{\V{q-p}'})
  \times[n_{\V{p}'}n_{\V{q}-\V{p}'}+(1-n_{\V{p}'})(1-n_{\V{q}-\V{p}'})]\hat{p}_{i}'\hat{p}_{j}'
\end{equation}
from \eqref{GG12} and \eqref{GG21}.
Let $n_{\V{p}'} = n^{\mathrm{eq}}(\eps_{\V{p}'}) + \delta n_{\V{p}'}$.
Recall that we want the first-order variation $\delta\Sigma$. 
\begin{multline}
  -\delta\big((GG)^{21}+(GG)^{12}\big)_{ij}=
  2\pi\int_{\mathbf{p}'}\delta(\omega-\epsilon_3-\epsilon_4) 
  \big[-\delta n_{3}-\delta n_{4}+2n_{3}\delta n_{4}+2n_{4}\delta n_{3}\big]\hat{p}_{i}'\hat{p}_{j}',
\end{multline}
where the subscripts $3$ and $4$ represent $\V{p}'$ and $\V{q}-\V{p}'$, and the superscript ``eq'' has been dropped.
After integrating the energy $\omega$, we have
\begin{multline}
\label{F-correction-1}
  \int\frac{d\omega}{2\pi}\,\frac{1}{2}\bigg(\frac{-\delta((GG)_{ij}^{21}+(GG)_{ij}^{12})}{|L_{\lambda}^{\mathrm{GL}}|^{2}}\bigg)\mathfrak{Re}G^{11} 
  =\frac{1}{2}\int_{\V{p}'}\frac{(2n_{4}-1)\delta n_{3}+(2n_{3}-1)\delta n_{4}}{|L_{\lambda}^{\mathrm{GL}}(\omega=\epsilon_{3}+\epsilon_{4},\mathbf{q})|^{2}}\times\frac{1}{\epsilon_{3}+\epsilon_{4}-\epsilon-\epsilon_{2}}\hat{p}_{i}'\hat{p}_{j}',
\end{multline}
where the external energy is $\eps$, and the energy $\eps_{2}\equiv \eps_{\V{q}-\V{p}}$ is for the internal fermion line in Figure~\refsub{fig-self-energies}{(d)}.

\begin{figure}[tbp!]
\centering
\begin{subfigure}[b]{0.49\textwidth}
\centering
\includegraphics[width=1.0\textwidth]{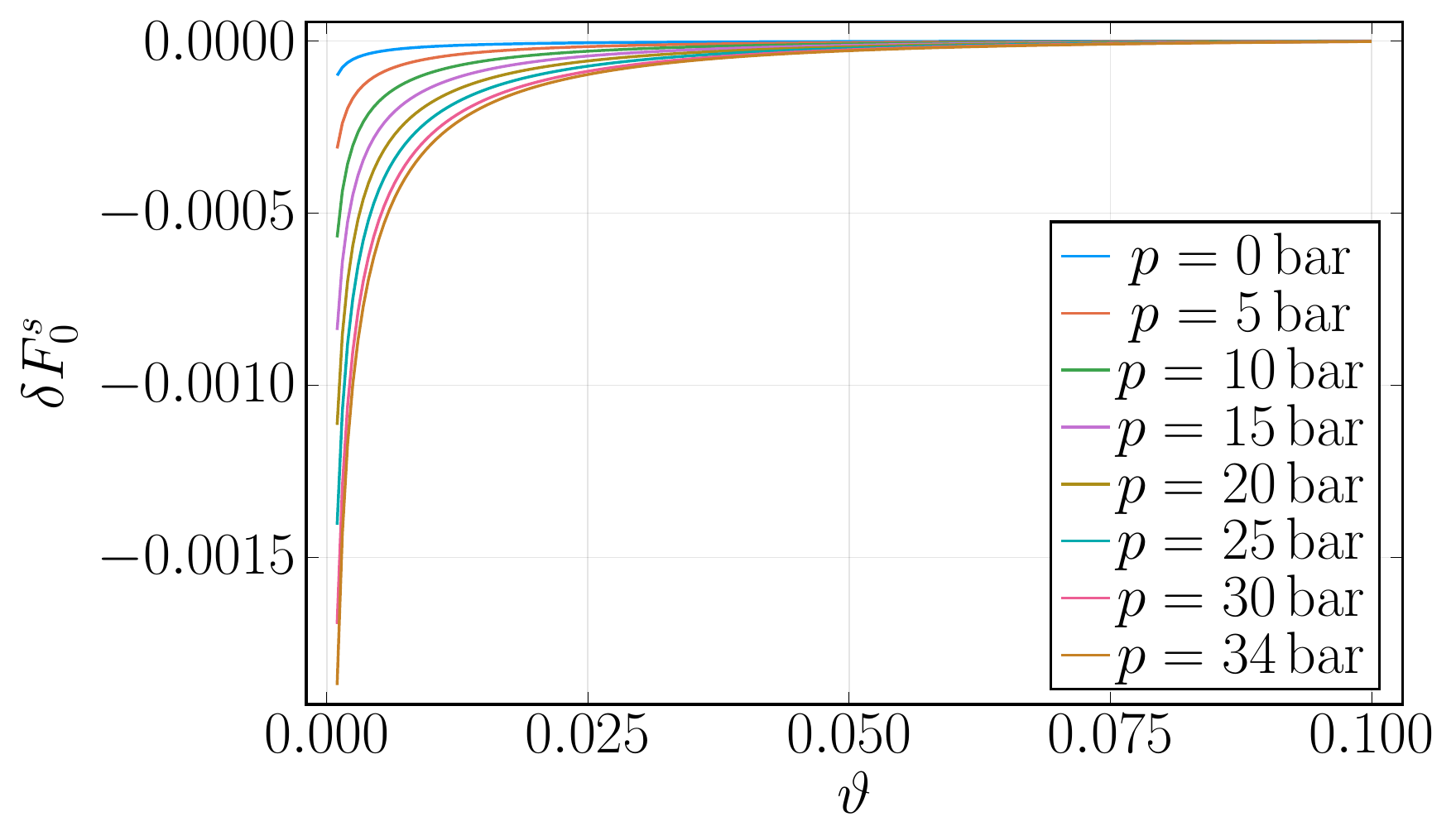}
\caption{}
\end{subfigure}
\hfill
\begin{subfigure}[b]{0.49\textwidth}
\centering
\includegraphics[width=1.0\textwidth]{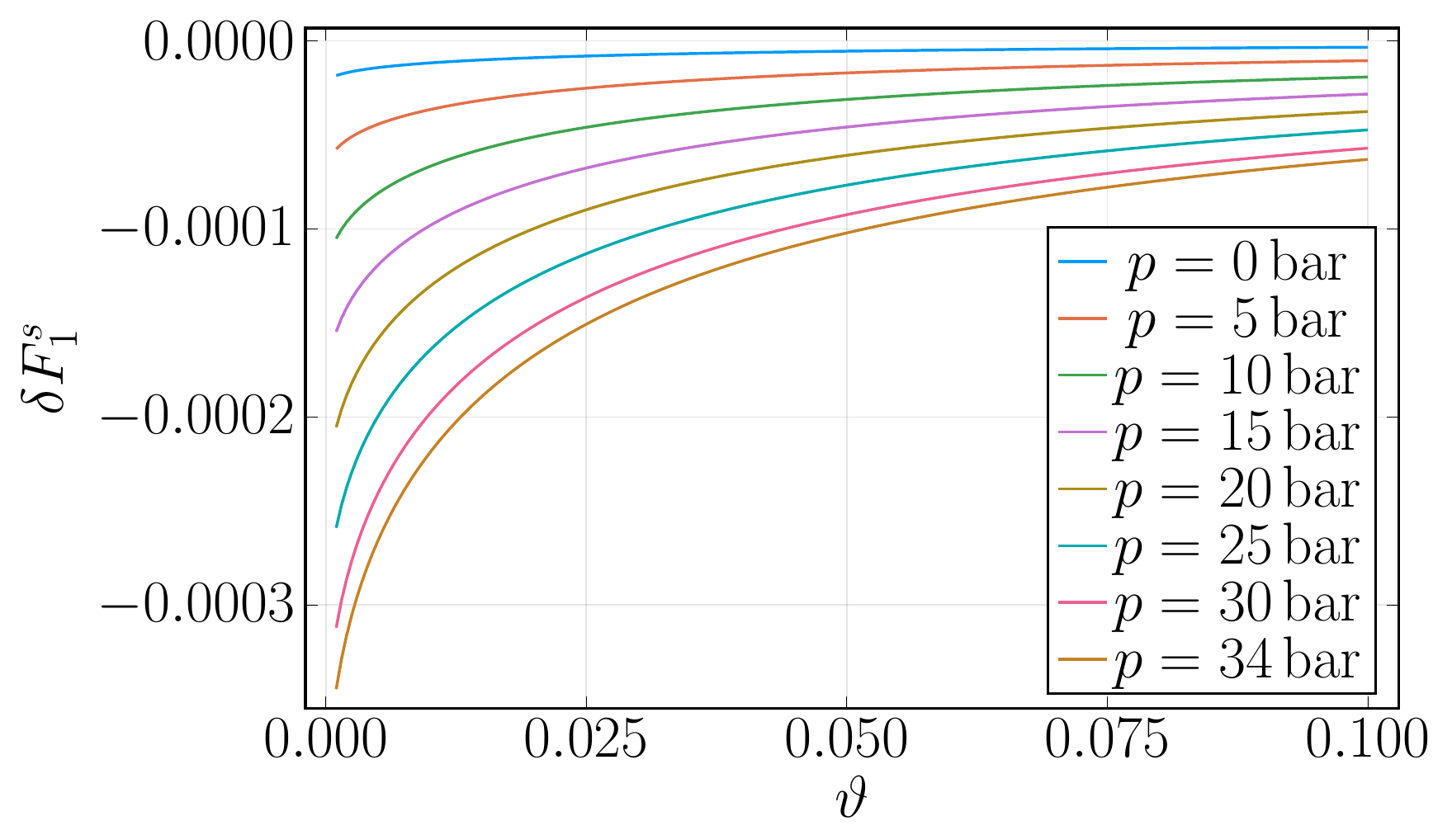}
\caption{}
\end{subfigure}
\caption[]{Panels (a) and (b) show the corrections to the Landau parameters $F^s_0$ and $F^s_1$, respectively, at different pressures as a function of $\vartheta$.
The curves are obtained from Eqs.~\eqref{F0} and \eqref{F1} and the pressure dependence of $1/N(0)\xi_0^3T_{\mathrm{c}}$.
The cutoff is $x_\mathrm{c}=0.236$.}
\label{fig-landau-param}
\end{figure}

For the second term in the right-hand side of \eqref{real-part}, the variation $\delta n$ can happen in both $(GG)$ terms and $\mathfrak{Im}G^{11}$.
From \eqref{LGL}, the variation gives
\begin{equation}
\delta((L_{ij}^{\mathrm{GL}})^{*}-L_{ij}^{\mathrm{GL}})=2i\int_{\mathbf{p}'}\frac{\delta n_{\V{p}'}+\delta n_{\V{q}-\V{p}'}}{\omega-\epsilon_{\mathbf{p'}}-\epsilon_{\mathbf{q-p'}}+i0}\hat{p}_{i}'\hat{p}_{j}'.
\end{equation}
Recall that
\begin{equation}
i\mathfrak{Im}G^{11}(q-p)=i\pi\delta(\omega-\epsilon-\epsilon_{2})[2n(q-p)-1].
\end{equation}
The $\omega$ integral gives
\begin{multline}
\label{F-correction-2}
  \int\frac{d\omega}{2\pi}\,\frac{1}{2}\bigg(\frac{\delta((L_{ij}^{\mathrm{GL}})^{*}-L_{ij}^{\mathrm{GL}})}{|L_{\lambda}^{\mathrm{GL}}|^{2}}\bigg)i\mathfrak{Im}G^{11} 
  =-\frac{1}{2}\int_{\V{p}'}\frac{(2n_{2}-1)}{|L_{\lambda}^{\mathrm{GL}}(\omega=\epsilon+\epsilon_{2},\mathbf{q})|^{2}}\times\frac{(\delta n_{3}+\delta n_{4})}{\epsilon+\epsilon_{2}-\epsilon_{3}-\epsilon_{4}}\hat{p}_{i}'\hat{p}_{j}'.
\end{multline}
For the $\mathfrak{Im}G^{11}$, 
\begin{equation}
i\delta\mathfrak{Im}G^{11}(q-p)=i\pi\delta(\omega-\epsilon-\epsilon_{2})\times2\delta n_{2}.
\end{equation}
In equilibrium, 
\begin{equation}
-L_{\lambda}^{\mathrm{GL}}+(L_{\lambda}^{\mathrm{GL}})^{*}=-2i\frac{N(0)}{3}(\vartheta+\xi_{\lambda}^{2}q^{2}).
\end{equation}
Thus the $\omega$ integral gives
\begin{equation}
\label{F-correction-3}
\int\frac{d\omega}{2\pi}\frac{1}{2}\frac{-L_{\lambda}^{\mathrm{GL}}+(L_{\lambda}^{\mathrm{GL}})^{*}}{|L_{\lambda}^{\mathrm{GL}}|^{2}}i\delta\mathfrak{Im}G^{11} \\
=\frac{1}{2}\frac{\frac{N(0)}{3}(\vartheta+\xi_{\lambda}^{2}q^{2})}{|L_{\lambda}^{\mathrm{GL}}(\omega=\epsilon+\epsilon_{2},\mathbf{q})|^{2}}\times2\delta n_{2}\times\hat{p}_{i}'P_{ij}^{\lambda}\hat{p}_{j}'.
\end{equation}
Note that the $\V{p}'$ integral has been done in this case. After the integration over $\V{q}$, the sum of \eqref{F-correction-1}, \eqref{F-correction-2} and \eqref{F-correction-3} gives the correction to the Landau parameters. The detail of the calculation can be found in the appendix. The results are the following.\footnote{We study the zero sound unpolarized in the spin space. Thus we only derive the corrections to the spin-symmetric Landau parameters $F^s_l$ here. The pair fluctuations should also influence spin-dependent properties, which are not considered in this article.}
\begin{equation}
\label{F0}
\begin{split}
\delta F_0^{s} =
-& 4.617\times10^{-3}\frac{1}{N(0)\xi_{0}^{3}T_{\mathrm{c}}}
\Big[\frac{\atanfactor}{\sqrt{\vartheta}}-\frac{x_{\mathrm{c}}}{x_{\mathrm{c}}^{2}+\vartheta}\Big]\\
+& 3.136\times10^{-2}\frac{1}{N(0)T_{\mathrm{c}}\xi_{0}^{3}}
\Big[x_{\mathrm{c}}-\frac{3}{2}\sqrt{\vartheta}\atanfactor)
+ \frac{\vartheta x_{\mathrm{c}}}{2(x_{\mathrm{c}}^{2}+\vartheta)}\Big] \\
+& 5.975\times10^{-2}\times\frac{1}{N(0)\xi_{0}^{3}T_{\mathrm{c}}}
\Big[x_{c}-\sqrt{\vartheta}\atanfactor\big)\Big]
\end{split}
\end{equation}
and
\begin{equation}
\label{F1}
\begin{split}
\delta F_1^{s} =
&1.211\times10^{-2}\frac{1}{N(0)T_{\mathrm{c}}\xi_{0}^{3}}
\Big[x_{\mathrm{c}}-\frac{3}{2}\sqrt{\vartheta}\atanfactor
+ \frac{\vartheta x_{\mathrm{c}}}{2(x_{\mathrm{c}}^{2}+\vartheta)}\Big] \\
-& 1.792\times10^{-1}\times\frac{1}{N(0)\xi_{0}^{3}T_{\mathrm{c}}}
\Big[x_{c}-\sqrt{\vartheta}\atanfactor\Big]
\end{split}
\end{equation}
The results are shown in Figure~\ref{fig-landau-param} at different pressures, using $x_{\mathrm{c}}=0.236$.
Note the Landau parameters in the normal state are $F^s_{0}\sim 100$ and $F^s_{1}\sim 10$, and thus the corrections from our result give $\delta F^s_i/F^s_i \lesssim 10^{-5}$ for both $i=0,1$, which is much smaller than the corrections to Fermi velocity and the time constant (see Figure~\ref{fig-fermi-velocity}).

As shown in Figures~\ref{fig-fermi-velocity} and \ref{fig-landau-param}, most Fermi liquid parameters have infrared divergence at $\vartheta=0$.
The perturbative calculation based on the normal-state fermion propagators is not valid when $\vartheta$ is too close to 0, because the correction would be large and strongly modify the normal-state propagator.
Nonetheless, if the temperature is not too close to $T_{\mathrm{c}}$, the corrections remain small and the perturbation method is valid.

\section{Conclusion}

Based on Keldysh's formulation of nonequilibrium field theory for Fermi systems we developed the theory for the leading-order corrections in the expansion parameter $\sml$ to the kinetic theory of Fermi liquids arising from the virtual emission and absorption of incipient Cooper pairs for temperatures above, but close to, a BCS pairing transition.
Key results for the quasiparticle-pair-fluctutuation contribution to the collision integral as well as corrections to the Fermi velocity and Fermi liquid interactions are reported. The theory is broadly applicable to interacting Fermi systems.   
As an application of the theory we calculated the effects of pair fluctuations on quasiparticle transport in liquid \He\ near the superfluid transition, including the excess attenuation of zero sound resulting from quasiparticle-pair-fluctuation scattering. The theory is in excellent agreement with the experimental results reported by Paulson and Wheatley, providing validation of our theory for the pair-fluctuation contribution to the quasiparticle collision integral.
We also report new results for the fluctuation corrections to the Fermi velocity and Fermi liquid parameters of liquid \He. In addition to these results, fluctuation corrections to other transport properties can be calculated from the theory presented here. We hope our theory motivates interest in pair-fluctuation studies in liquid \He\ as well as quasiparticle transport in unconventional superconductors.


\begin{acknowledgments}
  This work is supported by the National Science Foundation (Grant DMR-1508730).
\end{acknowledgments}

\appendix

\section{Correction to Landau Parameters}
In this appendix we derive Eqs.~\eqref{F0} and \eqref{F1} from Eq.~\eqref{F-correction-1}, \eqref{F-correction-2}, and \eqref{F-correction-3}.
The expression in Eq.~\eqref{F-correction-1} is symmetric under exchange of $\V{p}'$ and $\V{q}-\V{p}'$, so the expression can be reduced to
\begin{equation}
\int_{\V{q}}
\int_{\V{p}'}
\frac{(2n_{4}-1)\delta n_{3}}
{|L_{\lambda}^{\mathrm{GL}}(\omega=\epsilon_{3}+\epsilon_{4},\mathbf{q})|^{2}}
\times\frac{1}{\epsilon_{3}+\epsilon_{4}-\epsilon_1-\epsilon_{2}}\hat{p}_{i}'\hat{p}_{j}'.
\end{equation}
Moreover, we approximate
\begin{equation}
L_{\lambda}^\mathrm{GL}(\omega=\epsilon_{3}+\epsilon_{4},\mathbf{q})\approx
L_{\lambda}^\mathrm{GL}(\omega=0,\mathbf{q}).
\end{equation}
We will integrate the energy $\eps_{\V{p}'}$ and the momentum $\V{q}$.
The integral has a pole and we need to take the principal value.
\begin{equation}
\begin{split}
\frac{1}{\epsilon_{3}+\epsilon_{4}-\epsilon_{1}-\epsilon_{2}}
&=\frac{1}{\epsilon_{\mathbf{p}'}+\epsilon_{\mathbf{q}-\mathbf{p}'}-\epsilon_{\mathbf{p}}-\epsilon_{\mathbf{q}-\mathbf{p}}} \\
&\approx\frac{1}{2\epsilon_{\mathbf{p}'}-\mathbf{v}'\cdot\mathbf{q}-2\epsilon_{\mathbf{p}}+\mathbf{v}\cdot\mathbf{q}}.
\end{split}
\end{equation}
Changing variables,
\begin{equation}
\epsilon_{\mathbf{p}'}\to\epsilon_{\mathbf{p}'}+\epsilon_{\mathbf{p}}-(\mathbf{v}-\mathbf{v}')\cdot\mathbf{q}/2,
\end{equation}
the denominator becomes $2\eps_{\V{p}'}$.
We assume 
\begin{displaymath}
\delta n_{\V{p}'} = -\frac{\partial n_\mathrm{eq}}{\partial \eps_{\V{p}'}}\nu_{\hat{\V{p}}'}
\end{displaymath}
and $\delta n_{\V{p'+q}}\approx -n'(\eps_{\V{p'+q}})\nu_{\hat{\V{p}}'}$ so long as $q\ll p_\mathrm{F}$.
We neglect the distribution function $\nu_{\V{p}'}$ in the following for simplicity.
The numerator can be expanded around $\V{q}=0$, and we have
\begin{multline}
  \label{correction-1-integral}
\int_{\hat{\V{q}}}
  \int d\eps_{3}\frac{1}{2\eps_{3}}\Big[-(2n(\eps_{3})-1)n'(\eps_{3})
  -\frac{\V{q}\V{q}}{8}\big[(2f-1)f'''(\V{v}-\V{v}')(\V{v}-\V{v}') \\
  +4n'n''(\V{v}+\V{v}')(\V{v}-\V{v}') 
  +2n'n''(\V{v}+\V{v}')(\V{v}+\V{v}')\big]\Big]\hat{p}_{i}'\hat{p}_{j}',
\end{multline}
where all $n$'s and their derivatives are equilibrium distributions of the variable $\eps_{3}$,
and the first (second) vector $\V{q}$ is multiplied by the first (second) velocity vector.
Note that we have neglected the terms linear in $\V{q}$ since they vanish after the $\hat{\V{q}}$ integration.
Recall that we have a factor of $1/|L^{\mathrm{GL}}_{\lambda}|^2$.
Since the two components $\lambda=\parallel,\perp$ have different coherence lengths, we need to calculate them separately, by decomposing the integral \eqref{correction-1-integral} using $P_{ij}^{\lambda}$.
The rest is the radial $q$ integral, which can be done with a cutoff $q_{\mathrm{c}}$.
The final expression is a $\hat{\V{p}}'$ integral.
The Landau interaction energy can then be written as
\begin{equation}
\delta \eps_{\V{p}} = 
\int \frac{d\Omega_{\hat{\mathbf{p}}'}}{4\pi}
\sum_l F^s_l P_l(\hat{\mathbf{p}}\cdot\hat{\mathbf{p}}')
\nu_{\hat{\mathbf{p}}'}.
\end{equation}
Comparing the result with the above expression, we obtain the first two terms in \eqref{F0} and the first term in \eqref{F1}.

The integral \eqref{F-correction-2} does not contribute to the leading order.
The expansion gives
\begin{equation}
\int_{\V{q}} \int d\eps_3
\frac{1}{2\epsilon_{3}}\Big[-f'(\epsilon_{3})
-f''(\epsilon_{3})\big[-(\mathbf{v}-\mathbf{v}')\cdot\mathbf{q}/2\big] 
-\frac{1}{2}f'''(\epsilon_{3})\big[-(\mathbf{v}-\mathbf{v}')\cdot\mathbf{q}/2\big]^{2}\Big],
\end{equation}
where each term is either odd in $\eps_3$ or odd in $\V{q}$, so the whole integral vanishes.

The expression \eqref{F-correction-3} is
\begin{equation}
\frac{1}{2}\frac{\frac{N(0)}{3}(\vartheta+\xi_{\lambda}^{2}q^{2})}{|L_{\lambda}^{\mathrm{GL}}(\omega=\epsilon+\epsilon_{2},\mathbf{q})|^{2}}\times2\delta f_{2}\times\hat{p}_{i}'P_{ij}^{\lambda}\hat{p}_{j}'.
\end{equation}
Note that there is no $\mathbf{p}'$ integration for this term, which
has been done in obtaining the expression $N(0)(\vartheta+\xi_{0}^{2}q^{2})$.
Obviously, no $\hat{\mathbf{p}}\cdot\hat{\mathbf{p}}'$ term can appear
from this term. Integration over $\hat{\mathbf{q}}$ gives 
\begin{equation}
\label{A9}
\frac{1}{2}\frac{\frac{N(0)}{3}(\vartheta+\xi_{\lambda}^{2}q^{2})}{|L_{\lambda}^{\mathrm{GL}}(\omega=\epsilon+\epsilon_{2},\mathbf{q})|^{2}}\times2\delta f_{2}\times\frac{n_{\lambda}}{3}
\end{equation}
with $n_{\parallel}=1$ and $n_{\lambda}=2$.

Ignoring the $n_{\lambda}/3$ factor temporarily, the above expression \eqref{A9}
equals 
\begin{equation}
\frac{\frac{N(0)}{3}(\vartheta+\xi_{\lambda}^{2}q^{2})}{|L_{\lambda}^{\mathrm{GL}}(\omega=\epsilon+\epsilon_{2},\mathbf{q})|^{2}}\big[-f'(\epsilon_{\mathbf{q}-\mathbf{p}})\big]\nu_{-\hat{\mathbf{p}}}.
\end{equation}
Using the identity
\begin{equation}
\int\frac{d\Omega_{\hat{\mathbf{p}}'}}{4\pi}\Big[4\pi\sum_{lm}Y_{lm}(\hat{\mathbf{p}})Y_{lm}^{*}(\hat{\mathbf{p}}')\Big]g(\hat{\mathbf{p}}')=g(\hat{\mathbf{p}})
\end{equation}
for an arbitrary function $g$ defined on a sphere, the above expression
can be written as 
\begin{equation}
  \int\frac{d\Omega_{\hat{\mathbf{p}}'}}{4\pi}\frac{\frac{N(0)}{3}(\vartheta+\xi_{\lambda}^{2}q^{2})}{|L_{\lambda}^{\mathrm{GL}}(\omega=\epsilon+\epsilon_{2},\mathbf{q})|^{2}}\big[-f'(\epsilon_{\mathbf{q}-\mathbf{p}})\big] 
  \times\Big[4\pi\sum_{lm}Y_{lm}(\hat{\mathbf{p}})Y_{lm}^{*}(\hat{\mathbf{p}}')\Big]\nu_{-\hat{\mathbf{p}}'}.
\end{equation}
The addition theorem gives $(2l+1)P_{l}(\hat{\mathbf{p}}\cdot\hat{\mathbf{p}}')=4\pi\sum_{m}Y_{lm}(\hat{\mathbf{p}})Y_{lm}^{*}(\hat{\mathbf{p}}')$.
We then have 
\begin{equation}
  \int_{\hat{\mathbf{p}}'}\sum_{l}\frac{\frac{N(0)}{3}(\vartheta+\xi_{\lambda}^{2}q^{2})}{|L_{\lambda}^{\mathrm{GL}}(\omega=\epsilon+\epsilon_{2},\mathbf{q})|^{2}}\big[-f'(\epsilon_{\mathbf{q}-\mathbf{p}})\big] 
  \times(2l+1)(-1)^{l}P_{l}(\hat{\mathbf{p}}\cdot\hat{\mathbf{p}}')\nu_{\hat{\mathbf{p}}'},
\end{equation}
where the factor $(-1)^{l}$ comes from changing $-\hat{\mathbf{p}}'\to\hat{\mathbf{p}}'$.

Approximate $f'(\epsilon_{\mathbf{q}-\mathbf{p}})\approx f'(\epsilon_{\mathbf{p}})$
since $q$ only reduces the singularity. As before, we approximate
$\omega=\epsilon+\epsilon_{2}\approx0$ in the denominator. The $q$
integral gives
\begin{equation}
  \int\frac{q^{2}dq}{2\pi^{2}}\frac{1}{\frac{N(0)}{3}(\vartheta+\xi_{\lambda}^{2}q^{2})} 
  =\frac{1}{2\pi^{2}\frac{N(0)}{3}\xi_{\lambda}^{3}}\Big[x_{c}-\sqrt{\vartheta}\arctan\big(\frac{x_{c}}{\sqrt{\vartheta}}\big)\Big].
\end{equation}
Together with $-f'(\epsilon_{\mathbf{p}})=-f'(0)\approx1/4T$ in equilibrium,
we have 
\begin{equation}
  \int_{\hat{\mathbf{p}}'}\sum_{l}\frac{3}{8\pi^{2}N(0)\xi_{\lambda}^{3}T}\big[x_{c}-\sqrt{\vartheta}\arctan\big(\frac{x_{c}}{\sqrt{\vartheta}}\big)\big] 
  \times(2l+1)(-1)^{l}P_{l}(\hat{\mathbf{p}}\cdot\hat{\mathbf{p}}')\nu_{\hat{\mathbf{p}}'}.
\end{equation}
Using $\frac{1}{\xi_{\parallel}^{3}}+\frac{2}{\xi_{\perp}^{3}}\approx\frac{4.7174}{\xi_{0}^{3}}$, the final result gives the last term in \eqref{F0} and the last term in \eqref{F1}.

\section{Denominator of Equation~\eqref{alpha-0}}
\label{sec-denom}
In this appendix we simplify the denominator of Eq.~\eqref{alpha-0} to the result given in the expression~\eqref{denom}.
The denominator of \eqref{alpha-0} is 
\begin{equation}
\mathrm{Denom} = 
\sum_{l\ge 0}
\frac{(\nu_l^{(0)})^2}{2l+1}
\Big(1+\frac{F_l^{s}}{2l+1}\Big),
\end{equation}
which can be reduced to a simple expression since we assume $\nu_{l}=0$ for $l>2$.
From the equation for zero sound, we can obtain
\begin{equation}
\nu_1 = \frac{3s_0}{1+F^s_1/3}\nu_0
\end{equation}
and
\begin{equation}
\nu_2 = \frac{15}{2}\frac{s_0^2-s_1^2}{(1+F^s_1/3)(1+F^s_2/5)}\nu_0.
\end{equation}
See the article by Baym and Pethick for more details~\cite{bay78}.
The hydrodynamic sound velocity is given by~\cite{lan56}
\begin{equation}
s_1^2 = \frac{1}{3}(1+F^s_0)(1+F^s_1/3),
\end{equation}
where $s_1 \equiv c_1/v_{\fm}$,
and to a good approximation, the velocity difference between zero sound and first sound is given by~\cite{bay78}
\begin{equation}
s_0^2 - s_1^2 \approx \frac{4}{5}\frac{1+F^s_2/5}{1+F^s_0}s_1^2
= \frac{4}{15}\Big(1+\frac{F^s_1}{3}\Big)\Big(1+\frac{F^s_2}{5}\Big).
\end{equation}
Thus we have $\nu_{2} = 2\nu_{0}$.
Plugging these relations into the denominator, we have
\begin{equation}
\begin{split}
\mathrm{Denom} =\ & 
\nu_0^2 (1+F^s_0) +
\nu_0^2 \frac{3s_0^2}{1+F^s_1/3} +
\frac{4}{5}\nu_0^2\Big(1+\frac{F^s_2}{5}\Big) \\
=\ &
\frac{3\nu_0^2}{1+F^s_1/3}\bigg(\frac{1}{3}(1+F^s_0)\Big(1+\frac{F^s_1}{3}\Big) +
s_0^2 + \frac{4}{15}\Big(1+\frac{F^s_1}{3}\Big)\Big(1+\frac{F^s_2}{5}\Big)
\bigg) \\
=\ &
3\nu_0^2\frac{m}{m^*} 2s_0^2,
\end{split}
\end{equation}
where we have used the relation $\frac{m^{*}}{m}=1+\frac{F^s_{1}}{3}$~\cite{lan56}.


\end{document}